\documentclass[aps,pre,twocolumn,showpacs,superscriptaddress,amssymb]{revtex4}
\usepackage{graphicx,amsmath}
\newcommand{\outputfig}[3]
{\includegraphics[height=#2\linewidth,width=#3\linewidth]{#1}}

\begin{document}
\title{Semiclassical Study on Tunneling Processes via Complex-Domain Chaos}
\author{T. Onishi}
\email{t_onishi@comp.metro-u.ac.jp}
\affiliation{Department of Physics, Tokyo Metropolitan
University, Minami-Ohsawa, Hachioji 192-0397, Japan}
\author{A. Shudo} 
\affiliation{Department of Physics, Tokyo Metropolitan
University, Minami-Ohsawa, Hachioji 192-0397, Japan}
\author{K. S. Ikeda}
\affiliation{Faculty of Science and Engineering,
Ritsumeikan University, Noji-cho 1916, Kusatsu 525-0055, Japan}
\author{K. Takahashi}
\affiliation{The Physics Laboratories, Kyushu institute of
Technology, Kawazu 680-4, Iizuka 820-8502, Japan}
\date{\today}
\begin{abstract}
We investigate the semiclassical mechanism of tunneling process
in non-integrable systems.
The significant role of
complex-phase-space chaos
in the description of the tunneling process
is elucidated by studying a simple scattering map model.
Behaviors of tunneling orbits are encoded into symbolic sequences
based on the structure of complex homoclinic tanglement.
By means of the symbolic coding,
the phase space itineraries of tunneling orbits are related with
the amounts of imaginary parts of actions gained by the orbits,
so that the systematic search of significant tunneling orbits becomes
possible.
\end{abstract}
\maketitle

\section{\label{title_introduction}Introduction}

Tunneling is one of the most typical and important phenomena
in quantum physics, and for the past several years there is growing interest
in
natures of tunneling processes inherent in multi-dimensions.
Quantum properties in multi-dimensional systems have been investigated
extensively in terms of classical dynamical concepts
in the field of {\it quantum chaos}
\cite{Gutzwiller},
where the role of chaos, which is a generic  property
in multi-dimensional classical systems, has been elucidated.
It was found that tunneling properties are also strongly influenced
by whether underlying classical dynamics is chaotic or not
\cite{Wilkinson,Lin,Bohigas1,ShudoIkeda1,Frischat,Creagh,CreaghTunnelReview}
,
though tunneling process has no classical counterpart.

Tunneling occurs typically between classical invariant components
separated in phase space, such as between regular tori or chaotic seas.
On one hand, mechanism of tunneling
between distinct tori separated by chaotic seas has been studied
in the context of {\it chaos-assisted tunneling}
\cite{Bohigas1},
and its semiquantum analysis has been done, in which
the diffusion process in the chaotic sea accompanied with
tunneling paths from and into the tori around the boundaries of the sea
is considered to dominate the tunneling transport
\cite{Frischat}.
Experiments have also been performed by
measuring microwave spectra
in the superconducting cavity
\cite{Dembowski}, 
and measuring momentum distributions of cold atoms
in an amplitude-modulated standing wave of light
\cite{Hensinger_et_al,Steck_et_al}.

On the other hand, tunneling
between two chaotic seas separated by an energy barrier
has been studied by symmetric double wells
\cite{Creagh}.
It has been shown that the spectra of tunnel splittings
are reproduced by the orbits which consist of
instanton process under the barrier
and homoclinic exploration in each chaotic well.

Generic aspects of the link between tunneling process and
real domain process in non-integrable systems
have been examined in oscillatory scattering systems
\cite{TakahashiYoshimotoIkeda}.
They made an energy domain analysis for a model with continuous flow, 
while in the present study we make a time domain one for a scattering map. 
The semiclassical interpretation of complicated wave functions 
has been given in terms of oscillations of the stable manifold and 
an inherent property in flow systems, 
the divergent behavior of movable singularities of classical solutions 
on the complex time plane.

In near-integrable regime, the role of
resonances has been elucidated in the tunneling transport
between symmetric tori, by means of classical and quantum
perturbation theories
\cite{Brodier}.

In any case, if one wants to know mechanism of tunneling in chaotic
systems by relating it with the underlying classical structure, 
the use of complex orbits is inevitable
\cite{Miller},
since tunneling is a purely quantum mechanical process
and is not describable in terms of real classical
dynamics.  Full account of such a process should, therefore, be given
by {\it complex classical dynamics}.
An attempt to make a full complex semiclassical
analysis using the complex classical dynamics has been performed
to understand which kinds of complex trajectories describe
characteristic features of tunneling
in the presence of chaos, and how the complex classical dynamics
actually enters into real physical process
\cite{ShudoIkeda1,OnishiShudoIkedaTakahashi1,Adachi}.

In Ref.~\cite{ShudoIkeda1}, it was found that
the initial values of orbits which play a semiclassically primary role
form chain-like structures on an initial-value plane.
A phenomenology describing tunneling in the presence of chaos
based on such structures has been developed.

In Ref.~\cite{OnishiShudoIkedaTakahashi1},
the first evidence has been reported which
demonstrates the crucial role of {\it complex-phase-space chaos}
in the description of tunneling process
by analyzing a simple scattering map.
Also in this case, it was found that
initial values of orbits playing a semiclassically primary role
form chain-like structures on the initial-value plane.

Very recently, 
the chain-like structures are shown to be closely related to
the {\it Julia set} in complex dynamical systems
\cite{ShudoIshiiIkeda}.
The Julia set is defined as the boundary between
the orbits which diverge to infinity and
those which are bound for an indefinite time.
Chaos occurs only on the Julia set
\cite{Milnor}.
In Ref.~\cite{ShudoIshiiIkeda}, 
it was proven that
a class of orbits which potentially
contribute to a semiclassical wave function is identified as the Julia set.
It was also shown that
the transitivity of dynamics and high density of trajectories
on the Julia set characterize chaotic tunneling.

However, there still remains a problem
in complex semiclassical descriptions.
Significant tunneling orbits are always characterized by
a property that the amount of imaginary parts of classical actions
gained by the orbits are minimal among the whole candidates.
It is, however, difficult to find such significant orbits
out of the candidates,
because an exponential increase of the number of candidates
with time preventing us from
evaluating the amount of imaginary part of action
for every candidate.

To solve this problem, in this paper,
we investigate the structure of complex phase space
for a scattering map model,
and relate the structure
to the amounts of imaginary parts of actions
gained by tunneling orbits.
Our main idea is to relate the symbolic dynamics of
a homoclinic tanglement emerging in complex domain
to the behavior of tunneling orbits.
It enables us to estimate the amounts of imaginary parts of actions
gained by the tunneling orbits from symbolic sequences.

The organization of the paper is as follows.
In Sec. \ref{title_semiclssical_analysis_of_tunneling},
the symbolic description of tunneling orbits is developed. 
This description requires an effective symbolic dynamics 
constructed on a complex homoclinic tanglement. 
In that section, we emphasize the importance of the application 
of the symbolic dynamics to tunneling, 
and the details of how we construct the symbolic dynamics itself
is deferred to Sec. \ref{title_symbolic_dynamics}. 
So it should be noted that 
in Sec. \ref{title_semiclssical_analysis_of_tunneling} 
we use the result in Sec. \ref{title_symbolic_dynamics} 
without any technical details.

More precisely,
in Sec. \ref{title_semiclssical_analysis_of_tunneling},
the tunneling process is investigated by
the time-domain approach of complex semiclassical method.
We introduce a 
scattering map which would be the
simplest possible map modeling the energy barrier tunneling in
more than one degree of freedom.
Though real-domain chaos is absent in this model, it is shown that
tunneling wave functions exhibit the features
which have been observed in the systems creating real-domain chaos,
such as the existence of plateaus and cliffs in the tunneling
amplitudes and erratic oscillations on the plateaus.

It is elucidated that such tunneling features originate from
chaotic classical dynamics in the complex domain, in other words,
the emergence of homoclinic tanglement in the complex domain.
The symbolic description of the tanglement is introduced,
and is applied to the symbolic encoding of the behaviors
exhibited by semiclassical candidate orbits.
The amounts of imaginary parts of actions gained by the orbits
are estimated in terms of symbolic sequences assigned to the orbits.
Significant tunneling orbits are selected according to the estimation.

Finally, tunneling  wave functions are reproduced
in terms of such significant orbits, and
the characteristic features appearing in
tunneling amplitudes are explained by the interference
among such significant orbits.

In Sec. \ref{title_symbolic_dynamics},
the technical aspects which are skipped in 
Sec. \ref{title_semiclssical_analysis_of_tunneling}
are described in full details.
We first investigate the construction of a partition of complex phase space.
The homoclinic points are encoded into symbolic sequences
by means of 
the partition. 
Then some numerical observations are presented
which relates the symbolic sequences and
the locations of homoclinic points in phase space.
On the basis of the observations, we study
the relation between the symbolic sequences
and the amounts of imaginary parts of actions gained by
homoclinic orbits.
As a result, a symbolic formula
for the estimation of imaginary parts of actions is derived.

In Sec. \ref{title_discussion_and_conclusion},
we first conclude our present study, then
discuss the role of complex-domain chaos played
in semiclassical descriptions of tunneling
in non-integrable systems.
It is suggested that the complex-domain chaos
plays an important role in a wide range of
tunneling phenomena of non-integrable systems.
Finally some future problems are presented.

\section{Semiclassical Study on Tunneling Process via Complex-Domain Chaos}
\label{title_semiclssical_analysis_of_tunneling}

\subsection{Tunneling in a Simple Scattering System}
\label{title_simple_scattering_model}

We introduce a simple scattering map model which will be used in our study.
The Hamiltonian of our model is given as follows:
\begin{subequations}\label{hamiltonian}
\begin{eqnarray}
{\cal H}(q,p,t) & = & T(p) + V(q) \sum_{n=-\infty}^{+\infty}
\delta (t-n) ,                    \label{hamiltonian_a}\\
T(p) & = & p^2 /2 ,               \label{hamiltonian_b}\\
V(q) & = & k \exp (-\gamma q^2 ), \label{hamiltonian_c}
\end{eqnarray}
\end{subequations}
\noindent 
where $k$ and $\gamma$ are some parameters with positive values,
and the height and width of an energy barrier are given by
$k$ and $1/\sqrt{2\gamma}$ respectively.
A set of classical equations of motion is given as
\begin{subequations}\label{classical_equations}
\begin{eqnarray}
\left( q_{j+1}, p_{j+1} \right)
&=&  f 
\left( q_j, p_j \right), \\ \nonumber \\
 f :{\mathbb R}^2 \rightarrow {\mathbb R}^2\,\,\, \big| \,\,\,
\left( q, p \right) 
&\mapsto& 
\left( q + T^{\prime}(p), p - V^{\prime} (q+p) \right) , \ \ \
\end{eqnarray}
\end{subequations} 

\noindent
where $j$ denotes a time step with an integer value, and
the prime denotes a differentiation with respect to the corresponding
argument. 
${\mathbb R}^2$ denotes real phase space.

\begin{figure}
\outputfig{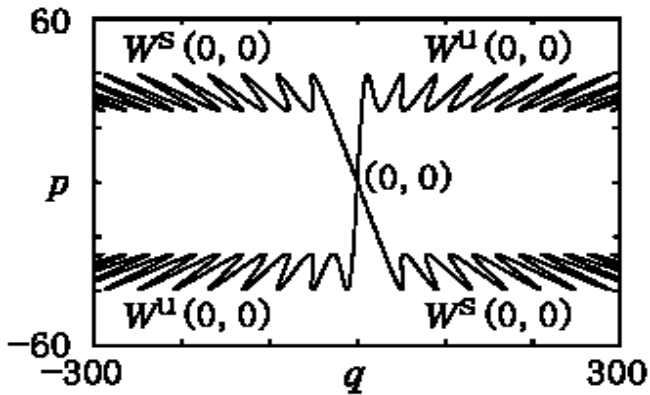}{0.6}{1.0}
\caption{\label{fig_real_domain_stb_unstb_mfds}
Real-domain stable and unstable manifolds of the unstable fixed point
located at the origin.
}
\end{figure}

\begin{figure}
\outputfig{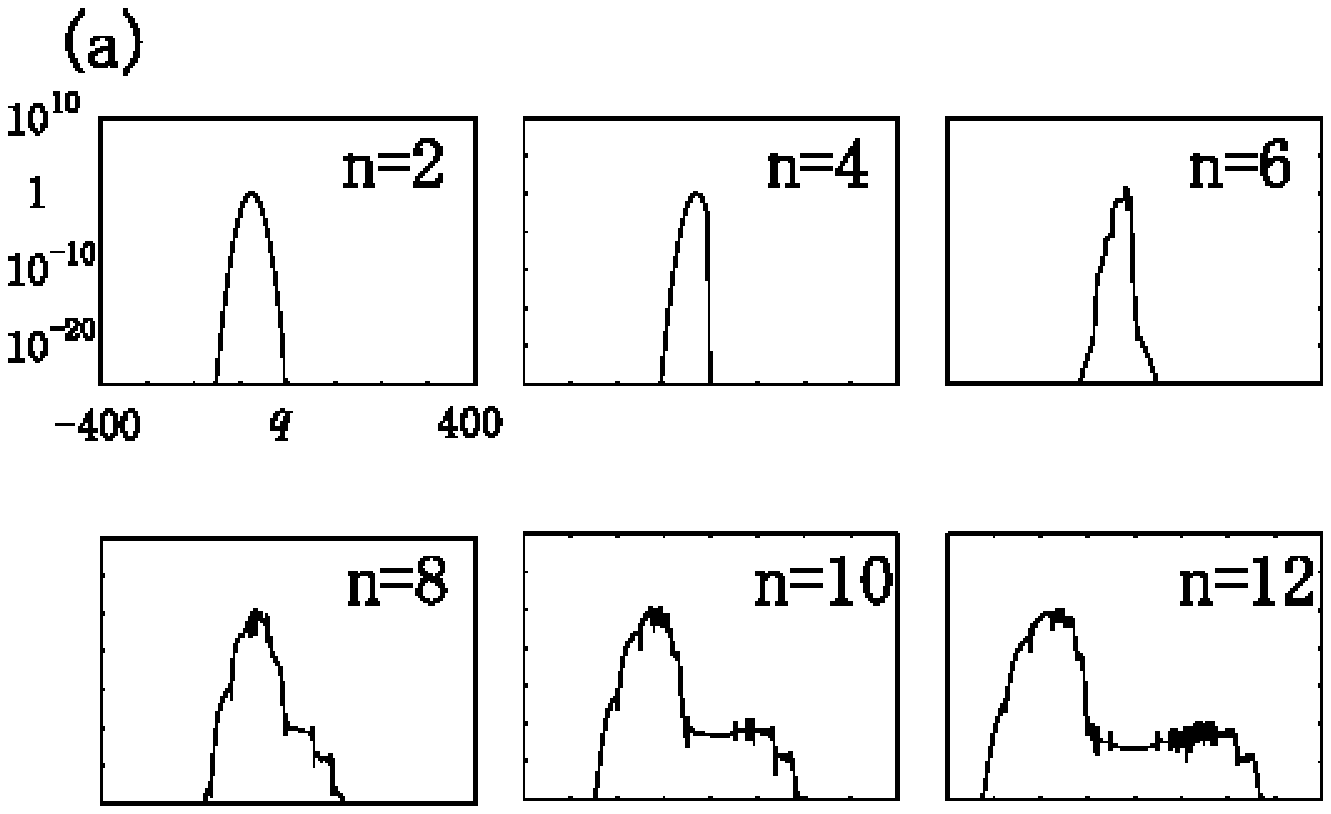}{0.56}{0.88}
\outputfig{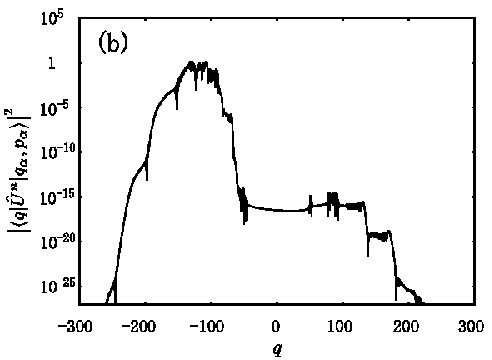}{0.6}{1.0}
\caption{\label{fig_quantum_time_evolution}
(a) Squared amplitudes of wave functions,
$\bigl| \langle q | \hat{U}^n | q_{\alpha}, p_{\alpha} \rangle \bigr| ^2$,
evaluated quantum-mechanically for $n=2$ to 12
in every 2 time steps
($\hbar \!=\! 1,$  $\sigma \!=\! 10,$ $k \!=\! 500, \gamma \!=\!
0.005,$  $q_{\alpha} \!=\! -123,$ $p_{\alpha} \!=\! 23$).
The incident wave packet is set
in the asymptotic region in the side of $q<0$.
In the lower figures, the center of mass has been already
reflected by the potential barrier located around the origin,
and those amplitudes observed
in the transmitted region represent tunneling effect.
(b) The enlarged figure at $n=10$.
}
\end{figure}

Fig.~\ref{fig_real_domain_stb_unstb_mfds} shows
the stable and unstable manifolds of a fixed point
located at the origin $(0,0)$.
These manifolds are denoted as ${\cal W}^s(0,0)$ and ${\cal W}^u(0,0)$
respectively. 
We should 
note that
the map does not create chaos in real phase space,
in contrast to the maps defined in bounded phase space
such as the standard map
\cite{LichtenbergLieberman}.
One can recognize this fact in several ways: for example,
the map has only single periodic orbit,
$(q,p)=(0,0)$. 
It means that the topological entropy of the system is null.
The other way is that
${\cal W}^s(0,0)$ and ${\cal W}^u(0,0)$
oscillate without creating homoclinic intersections.
Any classical manifold initially put on the real phase space is stretched
but not folded completely,
and it goes away to infinity along
${\cal W}^u(0,0)$.

The quantum mechanical propagation for a single time step
is given by the unitary operator:
\begin{equation}\label{unitary_operator}
\hat{U} = \exp\left[ -\frac{\rm i}{\hbar} V(\hat{q}) \right]
\exp\left[ -\frac{\rm i}{\hbar} T(\hat{p}) \right] ,
\end{equation}

\noindent
where $\hat{q}$ and $\hat{p}$ denote the quantum operators corresponding to
$q$ and $p$
respectively, 
which satisfy the uncertainty relation $[\hat{q},\hat{p}]\,=\,{\rm i}\hbar$.
A quantum incident wave packet is put in an asymptotic region.
The initial wave packet is given by a coherent state of the form:
\begin{equation}\label{incident_wave_packet}
\langle q | q_{\alpha}, p_{\alpha} \rangle
= \frac{1}{(\pi\hbar\sigma ^2 )^{1/4}}
\exp \left[ - \frac{ (q-q_{\alpha} )^2 }{ 2 \hbar \sigma ^2 }
-{\rm i} \frac{ p_{\alpha} (q_{\alpha} - 2q) }{ 2 \hbar } \right] ,
\end{equation}

\noindent
where $\sigma$ is a positive parameter and
the width of the wave packet in the $q$ direction is given by
$\sqrt{\hbar}\sigma$. 
$q_{\alpha}$ and $p_{\alpha}$ are the position and momentum of
the center of mass, respectively.
The initial kinetic energy is set to be far less than the potential barrier.

Fig.~\ref{fig_quantum_time_evolution} shows the iteration of 
the quantum wavepacket.  
Several features are observed
such as amplitude crossovers, the existence of plateaus and cliffs,
and erratic oscillations on the plateaus.
The same features 
have been reported in the case of dynamical tunneling
in mixed phase space
\cite{ShudoIkeda1}.
These are called the ``plateau-cliff structure'',
which has been confirmed in several systems
as a typical structure of tunneling wave functions
in the presence of real-domain chaos
\cite{ShudoIkeda1}.
However, as seen in our system,
the existence of the plateau-cliff structure
does not always need chaotic dynamics in real phase space.
So, 
the features of wave functions observed here
would be beyond
our intuitive expectation based on the real classical dynamics.
This strongly motivates the use of complex trajectories and
complex semiclassical analysis to describe
the features.

\subsection{Formulation of Semiclassical Analysis}
\label{title_formulation}

In the complex semiclassical analysis of our system,
it is convenient to define a pair of canonical variables 
$(Q,P)$ by 
\begin{subequations}\label{canonical_variables}
\begin{eqnarray}
Q & = & \frac{\sigma}{\sqrt{2}{\rm i}}(p+{\rm i}q \sigma ^{-2}), \\
P & = & \frac{\sigma}{\sqrt{2}}\,\,(p-{\rm i}q \sigma ^{-2}),
\end{eqnarray}
\end{subequations}

\noindent
and some notations by 
\begin{subequations}\label{some_notations}
\begin{eqnarray}
(Q_0, P_0) & = & \left( Q(q_0,p_0),P(q_0,p_0)\right), \\
(Q_{\alpha},P_{\alpha}) & = & 
\left( Q(q_{\alpha},p_{\alpha}), P(q_{\alpha},p_{\alpha})\right),
\end{eqnarray}
\end{subequations}
where $(q_0,p_0)$ and $(q_{\alpha},p_{\alpha})$ denote respectively
the initial value of the classical map $ f $, and
the value specifying the center of the wave packet
(\ref{incident_wave_packet}).

The wave function $\langle q | U^n | q_{\alpha}, p_{\alpha} \rangle$
is represented by an {\it n}-fold multiple integral:
\begin{eqnarray}
{\cal A}_n \int dq_0\ldots dq_{n-1}\exp \frac{\rm i}{\hbar}
\tilde{\cal S}_n,
\label{Feynman_integral}
\end{eqnarray}
which is a discrete analog of Feynman path integral, 
where 
\begin{subequations}
\begin{eqnarray}
& & {\cal A}_n = 
    (\pi\hbar\sigma^2 )^{-1/4} (2\pi {\rm i}\hbar )^{-n/2},
    \label{Van_Vleck_formula_b} \\
& & \tilde{S}_n = S_n + S_0 ,
    \label{Van_Vleck_formula_c} \\
& & S_n = \sum_{j=1}^{n} \left[ T(p_{j-1}) - V(q_j) \right] 
\quad (p_{j-1}= q_j - q_{j-1}) ,
    \label{Van_Vleck_formula_d} \\
& & S_0 = \frac{\rm i}{4}\left[ (Q_0 - Q_{\alpha} )^2 
    + (P_{\alpha} + {\rm i}Q_{\alpha})
    (P_{\alpha}+{\rm i}Q_{\alpha} -{\rm i}2Q_0 )\right], 
    \nonumber\\
& & \label{Van_Vleck_formula_e}   
\end{eqnarray}
\end{subequations}
A saddle point condition is imposed to the integral
to give the semiclassical Van Vleck's formula,
in which the wavefunction can be expressed by
purely classical-dynamical quantities.
The saddle point evaluation
with boundary conditions
yields 
a shooting problem joining the initial and final points,
$(q_0,p_0)$ and $(q_n,p_n)$.
In other words, 
the orbit should satisfy a set of classical equations of motion
and boundary conditions:
\begin{subequations}\label{shooting_problem}
\begin{eqnarray}
& & \left( q_{j+1}, p_{j+1} \right)
    =  f 
    \left( q_j, p_j \right) \quad (0\le j< n),
    \label{shooting_problem_b}\\
& & \left( q_0, p_0 \right) \in  {\cal I}, 
    \label{shooting_problem_c}\\
& & \left( q_n, p_n \right) \in  {\cal F},
    \label{shooting_problem_d}
\end{eqnarray}
\end{subequations}

\noindent
where $ f :{\mathbb C}^2\rightarrow {\mathbb C}^2$
is the classical map extended into complex
phase space,
and ${\cal I}$, ${\cal F}$ stand for manifolds defined by
\begin{subequations}\label{def_of_init_value_plane}
\begin{eqnarray}
{\cal I} & = & \{ (q,p)\in {\mathbb C}^2\,\,|\,\,
 P(q,p) = P_{\alpha} \} , \\
{\cal F} & = & \{ (q,p)\in {\mathbb C}^2\,\,|\,\,
\hbox{Im}\,q = 0\} .
\end{eqnarray}
\end{subequations}

We 
call the complex plane ${\cal I}$ the initial-value plane.
The final condition in (\ref{shooting_problem_d}) is
required  
since we here want to see  our wave function as a function of $q_n$,
which should take a real value.
Therefore, the solutions of
the classical equations of motion
are given by a subset of the initial-value plane:
\begin{eqnarray}\label{definition_of_M_set}
{\cal M}_n & = & {\cal I}\cap  f ^{-n}({\cal F} ).
\end{eqnarray}

\noindent
Since the initial 
``momentum''
$P_0$ is fixed as $P_{\alpha}$,
the shooting problem will be solved by adjusting
the initial ``position''
$Q_0$ in the initial-value plane ${\cal I}$.

The semiclassical Van Vleck's formula of 
the {\it n}-step wavefunction takes the form:
\begin{eqnarray}
\langle q_n | U^n |q_{\alpha}, p_{\alpha}\rangle 
&\approx & 
{\cal A}_n\!\!\!\!\!\! \sum_{(q_0,p_0)\in {\cal M}_n}
\left| 
\frac{\partial ^2 W_n}{\partial q_n \partial P_0 } \right|
^{\frac{1}{2}} \exp \frac{\rm i}{\hbar}
\left( \tilde{S}_n - \frac{\phi}{2} \right) , 
\nonumber \\
& & \label{Van_Vleck_formula_a}
\end{eqnarray}

\noindent
where the sum 
is over the complex orbits
whose initial points are located on ${\cal M}_n$
just defined.
$\phi(q_0,p_0)$ is the Maslov index of each complex orbit. 
$W_n(q_n,P_0)$ is a generating function which yields
a set of canonical transformations as
\begin{equation}\label{generating_function}
\left. \frac{\partial W_n}{\partial q_n}\right|_{P_0} = p_n , 
\text{ and }
\left. \frac{\partial W_n}{\partial P_0}\right|_{q_n} = -Q_0 .
\end{equation}

\noindent
The outline of the derivation of (\ref{Van_Vleck_formula_a}) follows
the conventional one
\cite{Tabor}.
Further details are also given in
the Appendix of Ref.~\cite{ShudoIkeda1}.

\subsection{Hierarchical Configuration of Initial Values}
\label{title_multi_generation_structure}

\begin{figure}
\outputfig{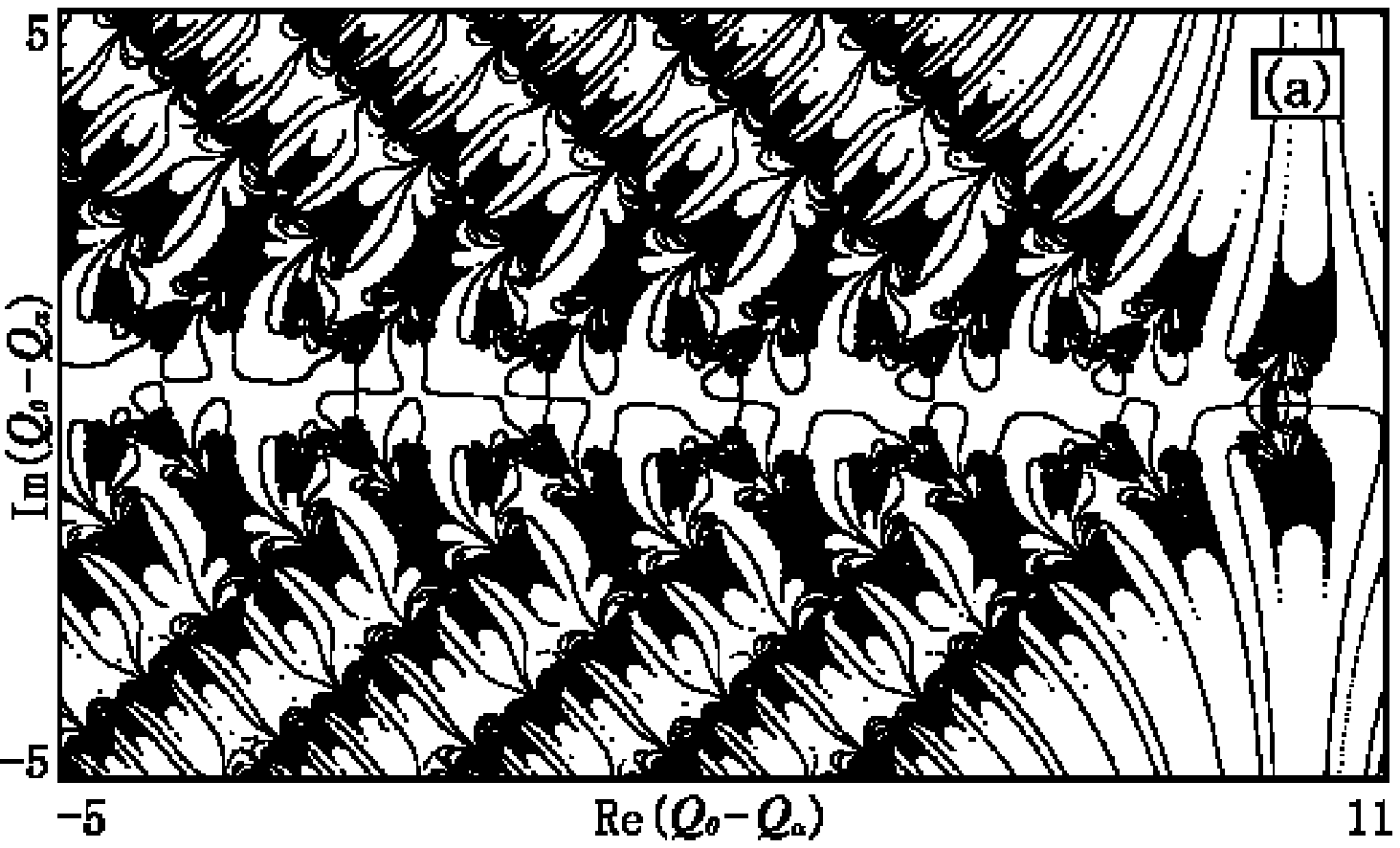}{0.7}{1.0}
\outputfig{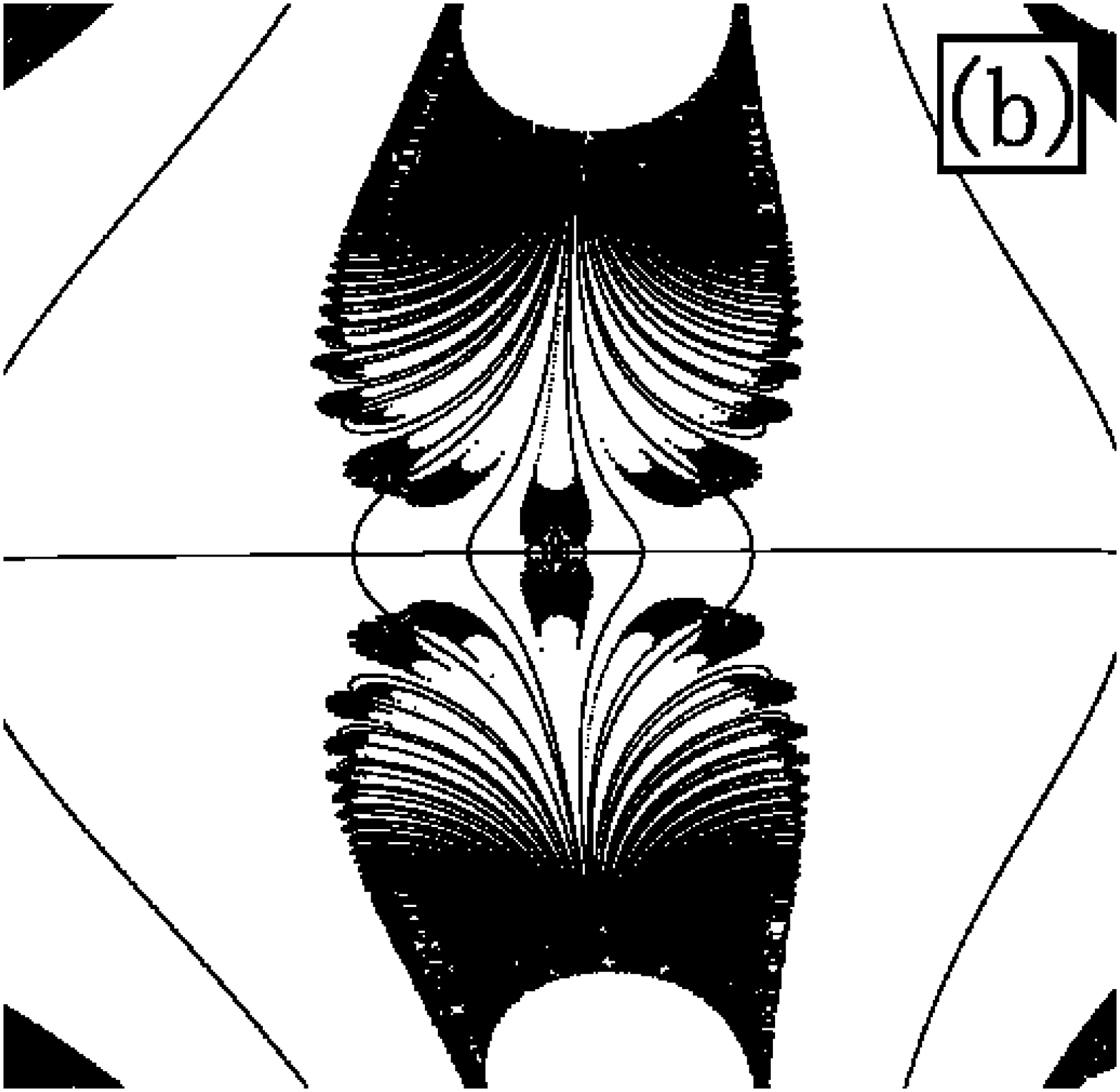}{0.6}{0.6}
\outputfig{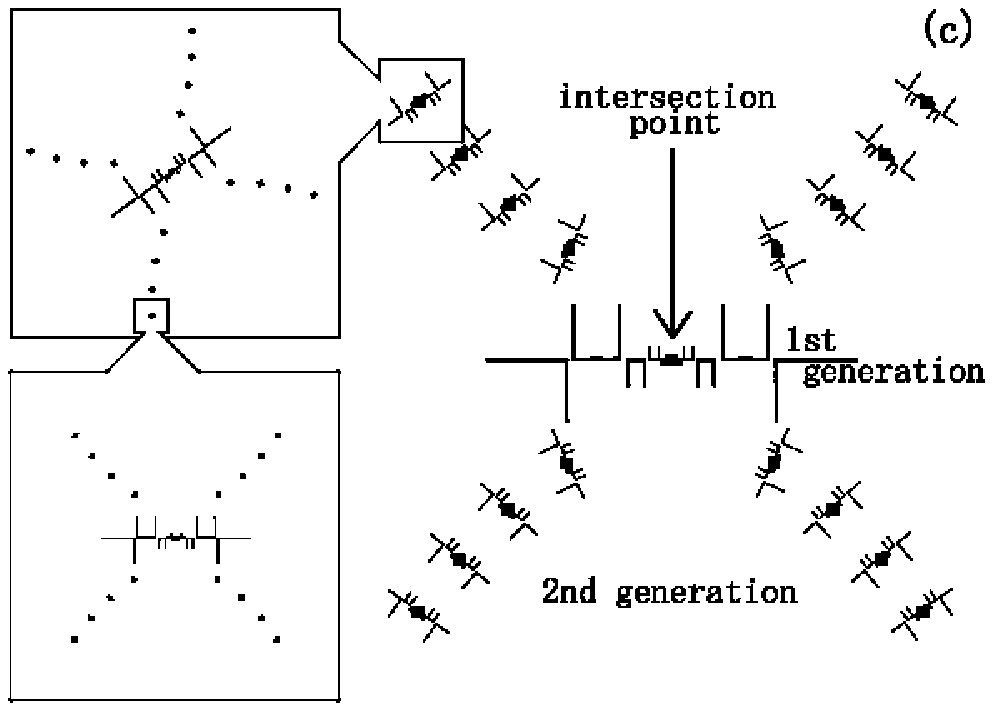}{0.6}{1.0}
\caption{\label{fig_M_set}
(a) ${\cal M}_n$ 
on the initial-value plane ${\cal I}$ for $n = 10$.
(b) The structure of ${\cal M}_n$
found in any small area on ${\cal I}$ where branches are densely aggregated.
(c) Schematic representation of (b).
Chain-like structures have a hierarchical configuration in ${\cal M}_n$.
A solid square located at the center of each chain-like structure
represents 
an element of 
${\cal I}\cap {\cal W}^s(0,0)$.
}
\end{figure}

As stated before, classical dynamics of the system
does not create real-domain chaos.
In contrast to that, the complex phase space has a complicated structure.
Fig.~\ref{fig_M_set}(a) shows a typical pattern of ${\cal M}_n$,
which consists of a huge number of strings.
A single string covers the whole range $(-\infty, +\infty)$
of the final $q_n$ axis, in other words,
each string has a solution of (\ref{shooting_problem})
for any choice of $q_n$.
So we call each string a branch.

The morphology of ${\cal M}_n$ is elusive.
To describe it, we introduce the notions of
chain-like structures and generations.
When  
blowing up any small area where branches are densely aggregated,
one can always find
the configuration of branches
as shown in Fig.~\ref{fig_M_set}(b).
The configuration is
drawn schematically in Fig.~\ref{fig_M_set}(c),
where a set of branches located in the center are linked to each other
in the horizontal direction with narrow gaps, and form
a {\it chain-like structure}
\cite{ShudoIkeda1,OnishiShudoIkedaTakahashi1}.
In any smaller area where branches are densely aggregated,
one can find again chain-like structures with finer scale.

As shown in Fig.~\ref{fig_M_set}(c),
small chain-like structures are arranged 
in both sides of the central large one, 
and the same arrangement repeats around each of the small chain-like
structures. 
This observation means that the branches in ${\cal M}_n$  
have a hierarchical configuration.  
Then it may be natural to assign a notion of {\it generation}
to each chain-like structure in the hierarchy.
For example, in Fig.~\ref{fig_M_set}(c),
we can say that the first 4 generations are displayed.

\begin{figure}
\outputfig{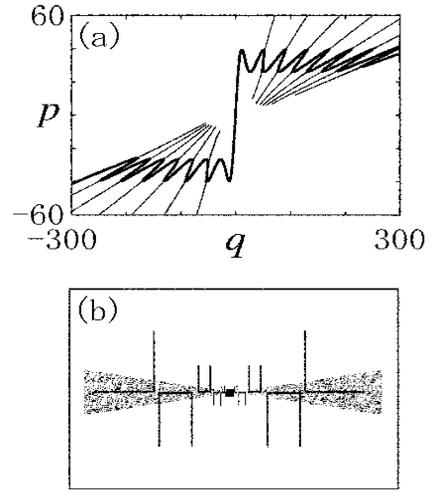}{0.77}{0.66}
\caption{\label{fig_M_and_L_sets}
(a) 
The $n$-step iteration of
${\cal M}_n\cap \Delta {\cal I}$ for $n=10$,
projected on real phase space.
Bold and thin parts almost agree with
${\cal W}^u(0,0)$ 
in real and complex phase spaces respectively 
(in the thin part, those points which have quite large 
$\hbox{Im}\,p_n$ are omitted).
(b) 
Schematic representation of
${\cal M}_n\cap \Delta {\cal I}$.
The center dot represents a point in 
$M$.
Hatched and non-hatched parts correspond to
the bold and thin parts in (a) respectively.
}
\end{figure}

\begin{figure}
\outputfig{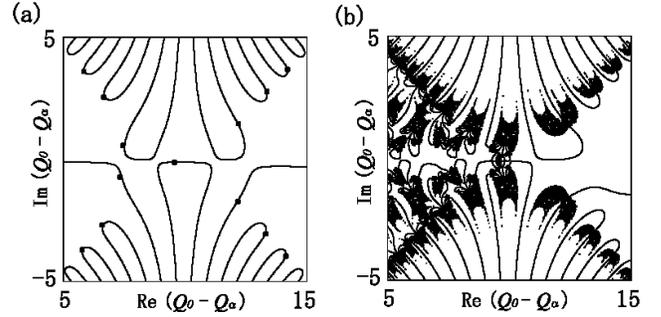}{0.5}{1.0}
\caption{\label{fig_creation_of_chain_around_intersection}
${\cal M}_n$ 
for (a) $n=2$ and (b) $n=10$,
in the same range of the plane ${\cal I}$.
In (a), the central five branches forms 
a chain-like structure, which is the only chain-like structure 
at $n=2$. 
Each filled square 
represents a point in
$M$. 
As $n$ increases, 
branches are densely aggregated 
in the neighbourhood of each filled square as shown in (b).
When enlarging the neighbourhood, 
one can always find 
the configuration of branches shown in Fig.~\ref{fig_M_set}(b).  
}
\end{figure}

\begin{figure}
\outputfig{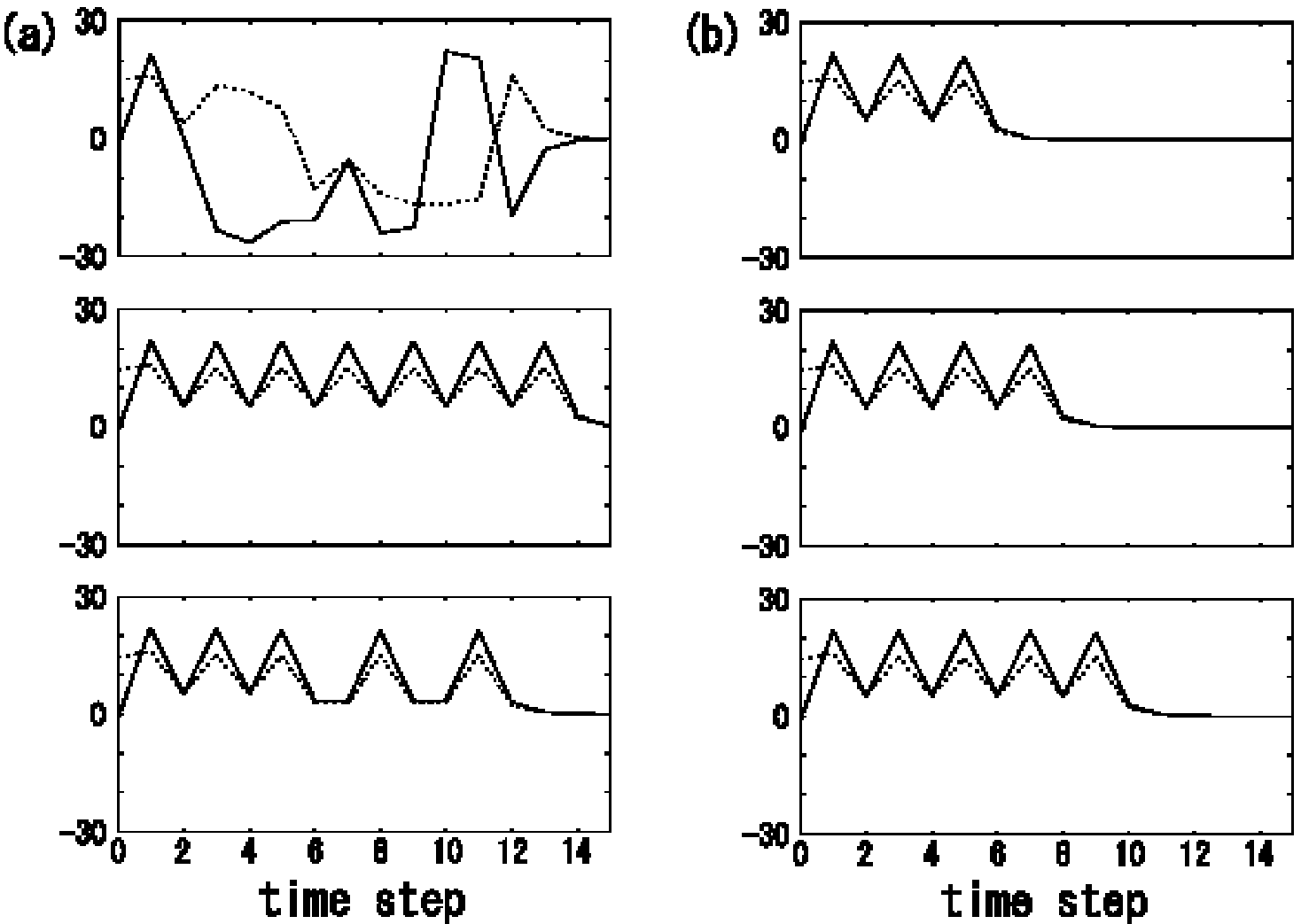}{0.6}{1.0}
\caption{\label{fig_orbits_in_MGS}
A variety of behaviors exhibited by
orbits launching from 
$M$. 
Solid and broken lines represent
$\hbox{Re}\,q$ and $\hbox{Im}\,q$ respectively.
In (b), 
the initial points of the trajectories belong to
the 6th, 8th, and 10th
generations respectively in the order from the top row to the bottom one.
}
\end{figure}

The hierarchical configuration of branches in ${\cal M}_n$
can be understood more clearly by relating it to
stable and unstable manifolds in complex 
domain.
We will show two facts:
first, 
chain-like structures are created by
the iteration of a small area element along
${\cal W}^s(0,0)$ and ${\cal W}^u(0,0)$.
Second, 
the intersection ${\cal I}\cap {\cal W}^s(0,0)$ forms the main frame
of the hierarchical configuration of branches.
In order to see these facts,
we use the notation:
\begin{eqnarray}
M & = & {\cal I}\cap {\cal W}^s(0,0).
\end{eqnarray}

\noindent
In the following, 
we mean a manifold with a real 1 dimension 
(resp. real 2 dimensions) by a ``curve'' (resp. ``surface'').

When the classical map $ f $ is extended
to the complex phase space ${\mathbb C}^2$,
both ${\cal W}^s(0,0)$ and ${\cal W}^u(0,0)$ 
are surfaces
in the complex domain
at least locally. 
Since the initial-value plane ${\cal I}$
is also the same dimensional surface,
the dimension of the intersection
$M$ is lower than 1 in general,
i.e., the intersection is neither a set of curves nor a surface,
but may be fractal like the Cantor set,
whose Hausdorf dimension may be between 0 and 1.
Let 
$(q_0,p_0)$ be an element of $M$,
and $\Delta {\cal I}$ be a
small neighbourhood of $(q_0,p_0)$
on the plane ${\cal I}$.
As the time proceeds, the orbit of
$(q_0,p_0)$
converges to 
the origin 
$(0,0)$ by definition.
On the other hand, the iteration of $\Delta {\cal I}$ approaches $(0,0)$
at the initial time stage, however, it is in turn spread
along 
${\cal W}^u(0,0)$,
and at a sufficiently large time step, it almost converges to ${\cal W}^u(0,0)$. 
We note that this process
is not specific to our system,
but the one that the stable and unstable manifolds always have.
Such a process takes place also in real-domain dynamics
or even in integrable one.

According to this mechanism and the conditions 
in (\ref{shooting_problem}),
one may expect that for a sufficiently large time step $n$,
the $n$-step iteration of 
${\cal M}_n\cap \Delta {\cal I}$ almost
agrees with a set:
\begin{eqnarray}\label{unstable_mfd_and_boundary_condition}
{\cal W}^u(0,0)\cap \{(q,p)\in {\mathbb C}^2
\,\,|\,\, \hbox{Im}\,q = 0\} .
\end{eqnarray}

\noindent
Even in the case of $n =10$,
the $n$-step iteration of ${\cal M}_n\cap \Delta {\cal I}$,
which is projected on the real phase space in 
Fig.~\ref{fig_M_and_L_sets}(a),
gives a sufficiently close manifold to that presented by
(\ref{unstable_mfd_and_boundary_condition}).

The bold and thin parts represent 
the parts of the final manifold which almost coincide with
the set (\ref{unstable_mfd_and_boundary_condition})
in real and complex phase spaces respectively.
This means that the bold part have sufficiently small
imaginary part of final momentum
compared to that of the thin part.
Fig.~\ref{fig_M_and_L_sets}(b) shows
$\Delta {\cal I}$ and
the schematic representation of ${\cal M}_n$.
The center dot represents
$(q_0,p_0)$
appearing in the above process.
Hatched part and non-hatched part correspond to
the bold and thin parts in Fig.~\ref{fig_M_and_L_sets}(a),
respectively.  Neighbouring branches are connected with each other
via caustics defined by $\partial q_n/\partial Q_0 |_{P_0} = 0$. 
The caustics are created by the oscillations of 
the real-domain ${\cal W}^u(0,0)$.

In this way, the creation of a chain-like structure can be understood by
considering the behavior of a small area element
first approaching real phase space with the guide of ${\cal W}^s(0,0)$,
then spreading over ${\cal W}^s(0,0)$.
In particular, the direction in which the real-domain ${\cal W}^u(0,0)$
stretches the area element determines
the direction of the chain-like structure on the plane ${\cal I}$.

It was found that a chain-like structure on the plane ${\cal I}$
is always created around a point in
$M$.
Fig.~\ref{fig_creation_of_chain_around_intersection} shows
the creation of chain-like structures around the points in $M$
as the time step proceeds.
In Fig.~\ref{fig_M_set}(c),
the points in 
$M$ 
are represented by filled squares.
That is to say, 
the hierarchical configuration of chain-like structures
implies those of the points in
$M$. 
Accordingly, 
$M$ constitutes the main frame of ${\cal M}_n$, and
the notion of generation is also assigned to each point in $M$.

Since, as mentioned above,
the iteration of initial points in chain-like structures
to real phase space
is described by orbits launching from $M$,
the structure of such orbits
will tell us 
that of orbits launching from ${\cal M}_n$,
the latter structure is
necessary for our
semiclassical analysis.
The study of orbits on the stable and unstable manifolds
is suitable for more canonical arguments since
they are compatible with
the theory of dynamical systems
\cite{Katok}.

Fig.~\ref{fig_orbits_in_MGS} shows
a variety of 
itineraries
of the orbits 
launching from 
$M$. 
In Fig.~\ref{fig_orbits_in_MGS}(a), the top row shows
a typical behavior observed in
$M$, where 
both $\hbox{Re}\,q$ and $\hbox{Im}\,q$
oscillate in an erratic manner for some initial time steps and
eventually approach the origin.
Regular itineraries such as periodic oscillations
coexist
among stochastic itineraries as shown in the middle row,
where an approximately periodic 2 behavior is
seen. 
Another type of orbit is shown in the bottom row,
where the trajectory first oscillates with period 2
and then turns into a periodic 3 motion.
The close relation between itinerating behaviors and
the notion of generation can be seen clearly
in the case of periodic oscillations, as shown in
Fig.~\ref{fig_orbits_in_MGS}(b).
In each row of the figure,
the 
period for which 
a trajectory oscillates
agrees with the generation of the initial point of the trajectory.

\subsection{Emergence of Homoclinic Tanglement in Complex Phase Space}
\label{title_emergence_of_homoclinic_tanglement}

Here we will show that the hierarchical structure of $M$,
which is the main frame of ${\cal M}_n$,
is the manifestation of complex-domain chaos.
First it is shown that
homoclinic tanglement emerges in the complex domain.
The laminations of ${\cal W}^s(0,0)$ and ${\cal W}^u(0,0)$
are densely developed around the origin $(0,0)$ in the complex domain,
and they are described by the notion of generation.
Then it is explained that the hierarchical structure of $M$
is created as a consequence of the emergence of
the complex homoclinic tanglement.

\begin{figure}
\outputfig{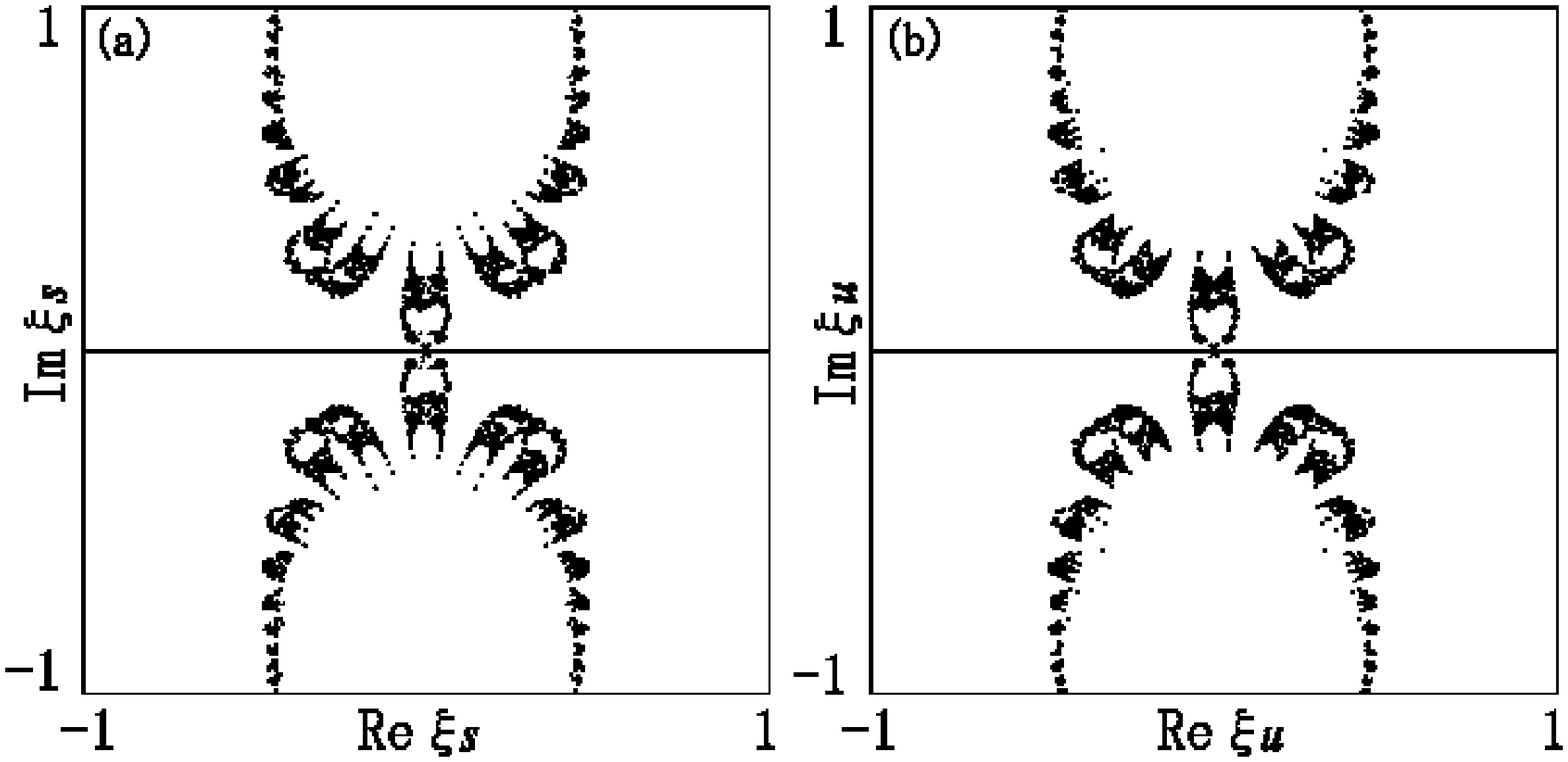}{0.5}{1.0}
\caption{\label{fig_homoclinic_pts}
The set of homoclinic points
associated with the unstable fixed point $(0,0)$
plotted on (a) ${\cal W}^s(0,0)$ and (b) ${\cal W}^u(0,0)$.
The origin of each figure corresponds to the unstable fixed point,
and the horizontal axis through the origin is included in real phase space.
}
\end{figure}

In order to study the structure of complex phase space,
we introduce coordinates on
${\cal W}^s(0,0)$ and ${\cal W}^u(0,0)$ as follows.
Let $\Phi_s$ and $\Phi_u$ be conjugation maps 
from a complex plane ${\mathbb C}$ 
to ${\cal W}^s(0,0)$ and ${\cal W}^u(0,0)$ respectively,
which satisfy the relations:
\begin{subequations}\label{def_of_normalized_coordinates}
\begin{eqnarray}
 f _s(\xi_s) & = & {\lambda}^{-1} \xi_s
\quad\text{ for }\xi_s \in {\mathbb C},
\label{def_of_normalized_coordinates_a} \\
 f _u(\xi_u) & = & \lambda \xi_u
\qquad\text{ for }\xi_u \in {\mathbb C},
\label{def_of_normalized_coordinates_b}
\end{eqnarray}
\end{subequations}

\noindent
where $ f _s$ and $ f _u$ are
\begin{subequations}
\begin{eqnarray}
 f _s(\xi ) & = & \left( \Phi_s^{-1}\,
 f \,\Phi_s \right) (\xi ), \\
 f _u(\xi ) & = & \left( \Phi_u^{-1}\,
 f \,\Phi_u \right) (\xi ),
\end{eqnarray}
\end{subequations}

\noindent
and $\lambda$ denotes
the maximal eigenvalue
of the tangent map on the unstable fixed point $(0,0)$
\cite{Gelfreich1}.
The above coordinates $\xi_s$ and $\xi_u$ are normalized in the sense that
the classical map $ f $
is represented by a linear transformation
on each coordinate.
The set of homoclinic points associated with the origin,
${\cal W}^s(0,0)\cap {\cal W}^u(0,0)$, is obtained numerically
on these coordinates
as shown in Fig.~\ref{fig_homoclinic_pts}.

In each of Fig.~\ref{fig_homoclinic_pts}(a) and (b),
enlarging the neighbourhood of the origin
by ${\lambda}^n$ times
$(n\ge 1)$,
one obtains 
the same figure with the original one
due to the relations in
(\ref{def_of_normalized_coordinates}).
This means that the homoclinic points are accumulated around
the origin $(0,0)$, in other words,
the laminations of ${\cal W}^u(0,0)$
and ${\cal W}^s(0,0)$ in the complex domain are densely developed around
the origin. 
It is a direct numerical evidence of the emergence of
{\it homoclinic tanglement} in complex phase space,
and thus 
null topological entropy in real phase space
does not always exclude the existence of chaos in complex domain.

\begin{figure}
\outputfig{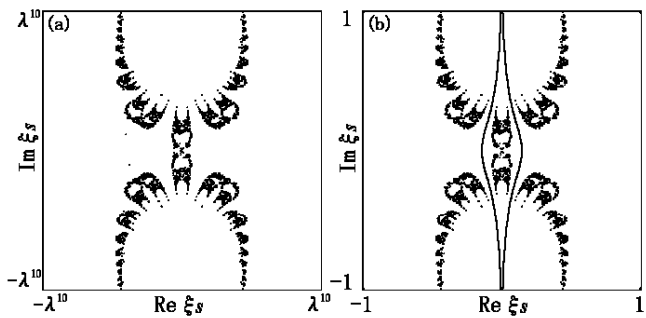}{0.5}{1.0}
\caption{\label{fig_domain_D_used_in_this_study}
(a) The intersection $M$ plotted on the $\xi_s$ plane.
(b) The domain $D$ plotted on the $\xi_s$ plane with
the set of homoclinic points superposed.
}
\end{figure}

Fig.~\ref{fig_domain_D_used_in_this_study}(a) shows
$M$ 
plotted on the $\xi_s$ coordinate.
Similarity to Fig.~\ref{fig_homoclinic_pts}(a) is evident,
which suggests that
the creation of 
the hierarchical configuration of $M$
on the plane ${\cal I}$
is due to the emergence of homoclinic tanglement in the complex domain.
The relation between
the structures of $M$ and the complex homoclinic tanglement
is made more clear by the notion of generation,
which we have already introduced.
To see the relation, the precise definition
of generation is given as follows.
Let $D$ be a connected domain in the $\xi_s$ plane satisfying
the conditions:
\begin{subequations}\label{conditions_of_domain_D}
\begin{eqnarray}
(0,0) & \in & D , \\
 f _s(D) & \subset & D ,
\end{eqnarray}
\end{subequations}

\noindent
where 
$ f _s$ is the linear transformation defined in
(\ref{def_of_normalized_coordinates_a}).
Then denoting $D\backslash  f _s(D)$ by $\tilde{D}$,
the $\xi_s$ plane is decomposed into a family
of disjoint domains as follows:
\begin{subequations}\label{def_of_generation}
\begin{eqnarray}
\bigsqcup_{n\in {\mathbb Z}}
 f _s^n(\tilde{D}) & = &
{\mathbb C} - \left\{ (0,0) \right\} ,
\label{def_of_generation_a}\\
 f _s^m(\tilde{D}) \cap
 f _s^n(\tilde{D}) & = & \phi
\quad (m \neq n) .
\label{def_of_generation_b}
\end{eqnarray}
\end{subequations}

\noindent
Thus for any point $\xi_s$ in this plane except $(0,0)$,
there exist a unique integer $n$ such that
$\xi_s\in  f _s^{-n}(\tilde{D})$.
We define the generation of the point $\xi_s$ as
such an integer $n$.

The shape of the domain $D$ is shown
in Fig.~\ref{fig_domain_D_used_in_this_study}(b).
Note that the relations in
(\ref{def_of_generation})
hold irrespective of the shape of
$D$. 
In our earlier publication
\cite{OnishiShudoIkedaTakahashi1},
$D$ was chosen as a disk.
In the present study,
another choice of $D$ is proposed.
More precise definition of $D$ is given in
Sec. \ref{title_symbolic_description_of_hierarchical_structure}
in the context of the construction of symbolic dynamics.

For any point in the neighbourhood of the origin in $D$,
the forward orbit approaches the origin straightforwardly
at an exponential rate
in the original coordinate $(q,p)$.
For any point in the $n$-th generation
$(n\ge 1)$, 
it takes at least $n$ steps until the orbit starts to approach the origin
exponentially, 
and thus it can exhibit a variety of behaviors
during its itinerary.
That is why the oscillations shown in
Fig.~\ref{fig_orbits_in_MGS}
are related with the generations.

\begin{figure}
\outputfig{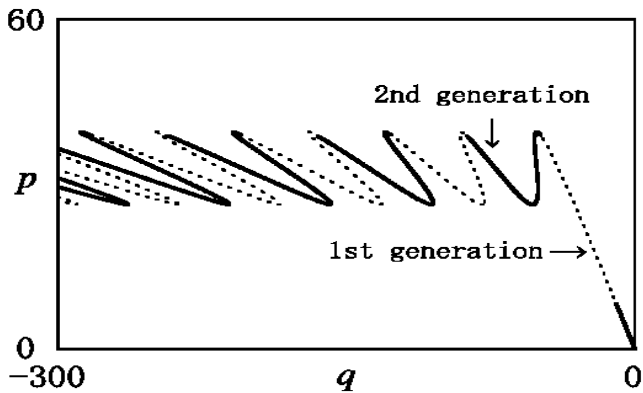}{0.6}{1.0}
\caption{\label{fig_generation_in_real_domain}
Generations of ${\cal W}^s(0,0)$ in real phase space.
The bold and dotted curves represent even and odd generations respectively,
except for the bold part emanating from the origin,
which represents the generations lower than or equal to zero.
The generation increases monotonically as $q\rightarrow -\infty$.
}
\end{figure}

Fig.~\ref{fig_generation_in_real_domain}
shows the generations in real phase space.
On one hand, 
in real-domain dynamics, a single iteration of the map $ f $
creates a single oscillation of ${\cal W}^s(0,0)$.
On the other hand, 
by definition, any point of ${\cal W}^s(0,0)$
in any generation is mapped by $ f $
to another point in the neighbouring generation.
Therefore in real phase space,
each generation corresponds to a single oscillation of ${\cal W}^s(0,0)$.

In the complex domain, however,
as seen in  Fig.~\ref{fig_M_set}(c),
higher generations describe finer scales of
the hierarchical configuration of $M$
on the plane ${\cal I}$.
Fig.~\ref{fig_M_set}(c) also shows that
for any point of $M$ and
in its small neighbourhood on the plane ${\cal I}$,
the number of the members
in $M$ which are included in individual generations
increases exponentially with
the generations.  
This situation clearly shows that 
the homoclinic tanglement of 
${\cal W}^s(0,0)$ and ${\cal W}^u(0,0)$ creates
the hierarchical configuration of $M$ on the plane ${\cal I}$.

\begin{figure}
\outputfig{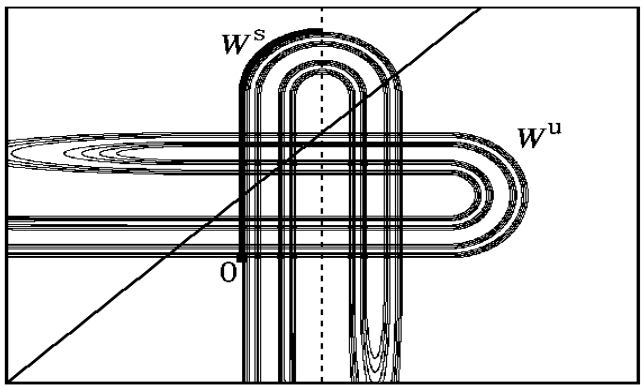}{0.6}{1.0}
\caption{\label{fig_analogy_with_hoese_shoe_map}
The horse-shoe map defined on a two-dimensional plane.
The one-dimensional bar across the tanglement corresponds to
our initial-value plane ${\cal I}$,
and the bold curve on ${\cal W}^s(0,0)$ corresponds to our domain $D$.
The dotted line is the boundary of partition which creates binary codes.
}
\end{figure}

It should be noted that real-domain chaos, if it exits, also creates
the hierarchical configuration.
Fig.~\ref{fig_analogy_with_hoese_shoe_map} shows
a simple analogy of our present situation
with a horse shoe map on a two-dimensional plane.
One-dimensional ${\cal W}^s(0,0)$ and ${\cal W}^u(0,0)$
create homoclinic tanglement,
and the one-dimensional bar across the tanglement
corresponds to our initial-value plane ${\cal I}$.
In this case, both ${\cal I}\cap {\cal W}^s(0,0)$
and ${\cal I}\cap {\cal W}^u(0,0)$ form the Cantor sets,
and the fractal structures of these intersections
are originated from densely developed laminations
of ${\cal W}^s(0,0)$ and ${\cal W}^u(0,0)$.
The 
interval on ${\cal W}^s(0,0)$ indicated by
a bold curve corresponds to our domain $D$.
Replacing ${\mathbb C}$ in (\ref{def_of_generation_a}) 
by ${\mathbb R}$,
one can define generations in a similar way.
Due to the horse shoe dynamics on this plane, 
${\cal W}^s(0,0)$ in the $n$-th generation has
$2^n$ intersection points with the initial-value bar.

In this sense, 
whether in real domain or in complex one,
the hierarchical configuration
of ${\cal I}\cap {\cal W}^s(0,0)$
on the plane ${\cal I}$
is nothing but the manifestation of chaos.
In particular, the emergence of
the above configuration
in the complex domain is a piece of evidence for
complex-domain chaos.
Once we know that chaos exists also in the complex
domain,
the methodology studying chaos in the real
domain 
can be applied to the analysis on the complex
domain.   
In particular,
symbolic dynamical description of
orbits, 
which is available if one finds a proper partition
of phase space 
to define it,
is a standard technique in the theory of dynamical systems
\cite{Katok}, 
and can be very useful tool to analyze complicated
phase space structures.  Homoclinic orbits are also describable
in terms of the symbolic dynamics, so our strategy
to study 
the hierarchical configuration of $M$
hereafter 
is to take the symbolic description of homoclinic orbits.

\subsection{Symbolic Description of Complex Orbits}
\label{title_symbolic_description_of_hierarchical_structure}

We will explain the symbolic description of
the complex homoclinic orbits and its application to
the symbolic description of semiclassical candidate orbits.
The symbolic description of homoclinic orbits contains
the construction of symbolic dynamics which works effectively and
the estimation of imaginary parts of actions for the homoclinic orbits.
We here present only the final results of our study on this description,
and the detailed explanation is given in
Sec. \ref{title_symbolic_dynamics}.
On the basis of the results,
the symbolic description of semiclassical candidate orbits
is discussed as follows.
First the orbits launching at $M\,(={\cal I}\cap {\cal W}^s(0,0))$
are encoded into symbolic sequences.
We compare both configurations of homoclinic points 
and the elements of $M$, on the $\xi_s$ coordinate 
set on ${\cal W}^s(0,0)$. 
A clear similarity between both configurations 
enables us to find, for each element of $M$, 
a homoclinic point located in the neighbourhood 
of the element of $M$. 
Since the homoclinic point is encoded into a symbolic sequence, 
we encode the element of $M$ into this symbolic sequence.  
Then semiclassical candidate orbits are encoded into symbolic sequences.
Since the behaviors of orbits launching at a single chain-like structure
are described by the behavior of the orbit of an element of $M$ located
at the center of the chain-like structure,
we assign the symbolic sequence of the element of $M$
to the initial points of the chain-like structure.

Symbolic dynamics is usually constructed by
finding a generating partition ${\cal G}$,
which is the partition of phase space
satisfying the relation:
\begin{eqnarray}\label{def_generating_partition}
\bigvee_{n=-\infty}^{+\infty} f ^n({\cal G}) &=& \epsilon_0,
\end{eqnarray}

\noindent
where the l.h.s. is the product of all partitions created by
the iterations of ${\cal G}$,
and the r.h.s. is the partition of phase space into
its individual points
\cite{Katok} 
(here we should consider the ``phase space''
as the closure of a set of homoclinic points).
Roughly speaking, ${\cal G}$ is the partition of phase space such that
each separated component of phase space corresponds to
a symbol and for every bi-infinite sequence of symbols
there may at most exist one trajectory of the
original map.

In our case, a partition of complex phase space
is defined in terms of 
a phase part of the gradient of the potential function:
\begin{eqnarray}\label{phase_of_gradient}
V^{\prime}(q) &=& -2\gamma k \exp\left[ -A(q)- {\rm i}B(q) \right] ,
\end{eqnarray}

\noindent
where $A(q), B(q)\in {\mathbb R}$.
The boundaries of our partition
are defined as follows:
\begin{eqnarray}
\left\{ (q,p)\in {\mathbb C}^2\,\,|\,\, B(q)=
\left[2\nu xy -(3x+1)y /2\right] \pi \right\} ,
\label{def_of_boundary_in_Sec_II}
\end{eqnarray}

\noindent
where $\nu$ is 
arbitrary 
positive integer,
and
$(x,y)$ is an element of the set:
\begin{eqnarray}\label{def_pairs_of_integers}
{\cal T} & = & \left\{ (+1,+1),(-1,+1),(-1,-1),(+1,-1)\right\} .
\end{eqnarray}

\begin{figure}
\outputfig{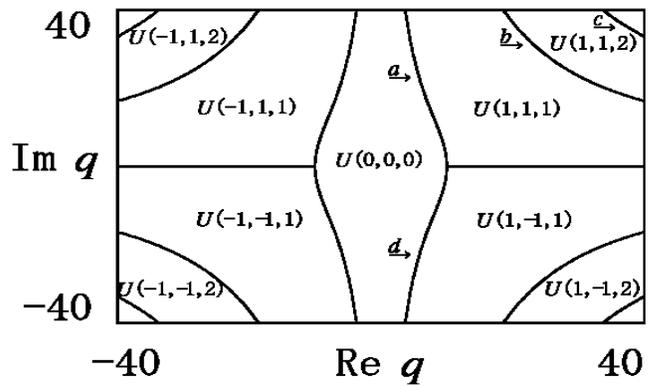}{0.6}{1.0}
\caption{\label{fig_mfd_C}
Intersection between the $q$ plane and the boundaries of the partition.
The curves labeled by $a,b,c,$ and $d$ indicate the boundaries
associated with $(x,y,\nu)=(1,1,1), (1,1,2), (1,1,3),$ and $(1,-1,1)$
respectively. 
}
\end{figure}

\noindent
The intersection between the $q$ plane and
the boundaries of the partition
are shown in Fig.~\ref{fig_mfd_C}.
Each boundary is a three-dimensional manifold
according to (\ref{def_of_boundary_in_Sec_II}),
so that the intersection is a set of one-dimensional curves.
$x$ and $y$ in (\ref{def_of_boundary_in_Sec_II})
represent the signs of $\hbox{Re}\,q$ and $\hbox{Im}\,q$
of the points on a boundary respectively,
i.e., the pair of $x$ and $y$ specifies the quadrant
of the $q$ plane where the $q$ component of the boundary is included.
$\nu$ is the ``distance'' between the boundary and the origin $(0,0)$, 
in the sense that the axis $\hbox{Re}\,q=\hbox{Im}\,q$ 
on the $q$ plane intersects the boundary at 
$q\approx (x\sqrt{\nu\pi /\gamma},y\sqrt{\nu\pi /\gamma})$ 
for $\nu\gg 1$.

Then our partition, denoted by ${\cal P}$, is defined
as a set of phase space components as follows:
\begin{subequations}\label{def_partition_and_symbolic_space}
\begin{eqnarray}
{\cal P} & = & \left\{ U(x,y,\nu )\,\,|\,\,
(x,y,\nu )\in {\cal S} \right\},
\label{def_of_partition}\\
{\cal S} & = & ({\cal T}\otimes {\mathbb N})\cup\{ (0,0,0)\} ,
\label{def_of_symbols}
\end{eqnarray}
\end{subequations}

\noindent
where 
$U(x,y,\nu )$ for $(x,y,\nu )\in {\cal T}\otimes {\mathbb N}$,
whose $q$ component is displayed in Fig.~\ref{fig_mfd_C}, denotes
the region enclosed by two boundaries
associated with $(x,y,\nu )$ and $(x,y,\nu +1)$,
and $U(0,0,0)$ denotes
the complement of the union of all components associated with
$(x,y,\nu)\in {\cal T}\otimes {\mathbb N}$.
The origin of phase space, $(0,0)$, is
included
in $U(0,0,0)$.

\begin{figure}
\outputfig{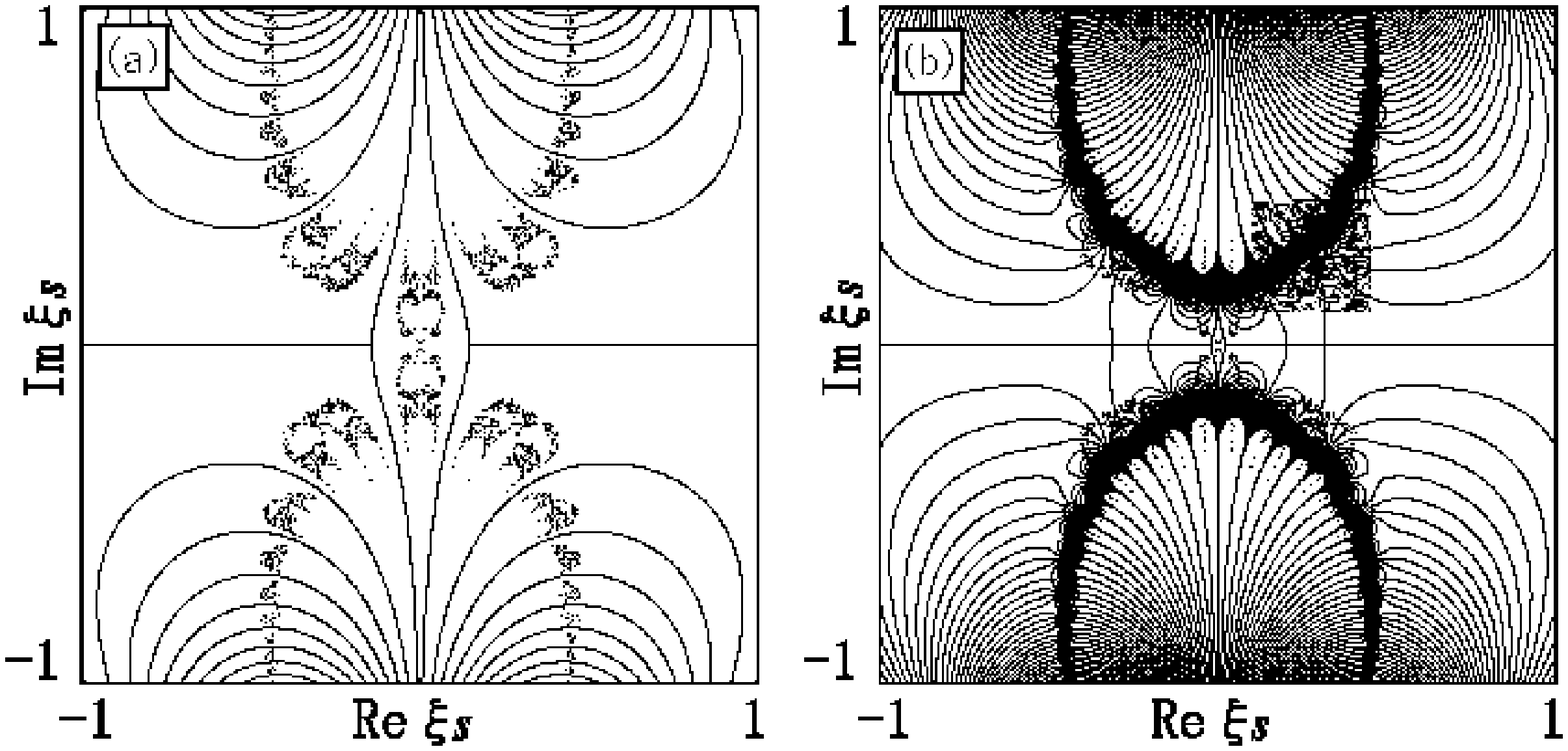}{0.5}{1.0}
\outputfig{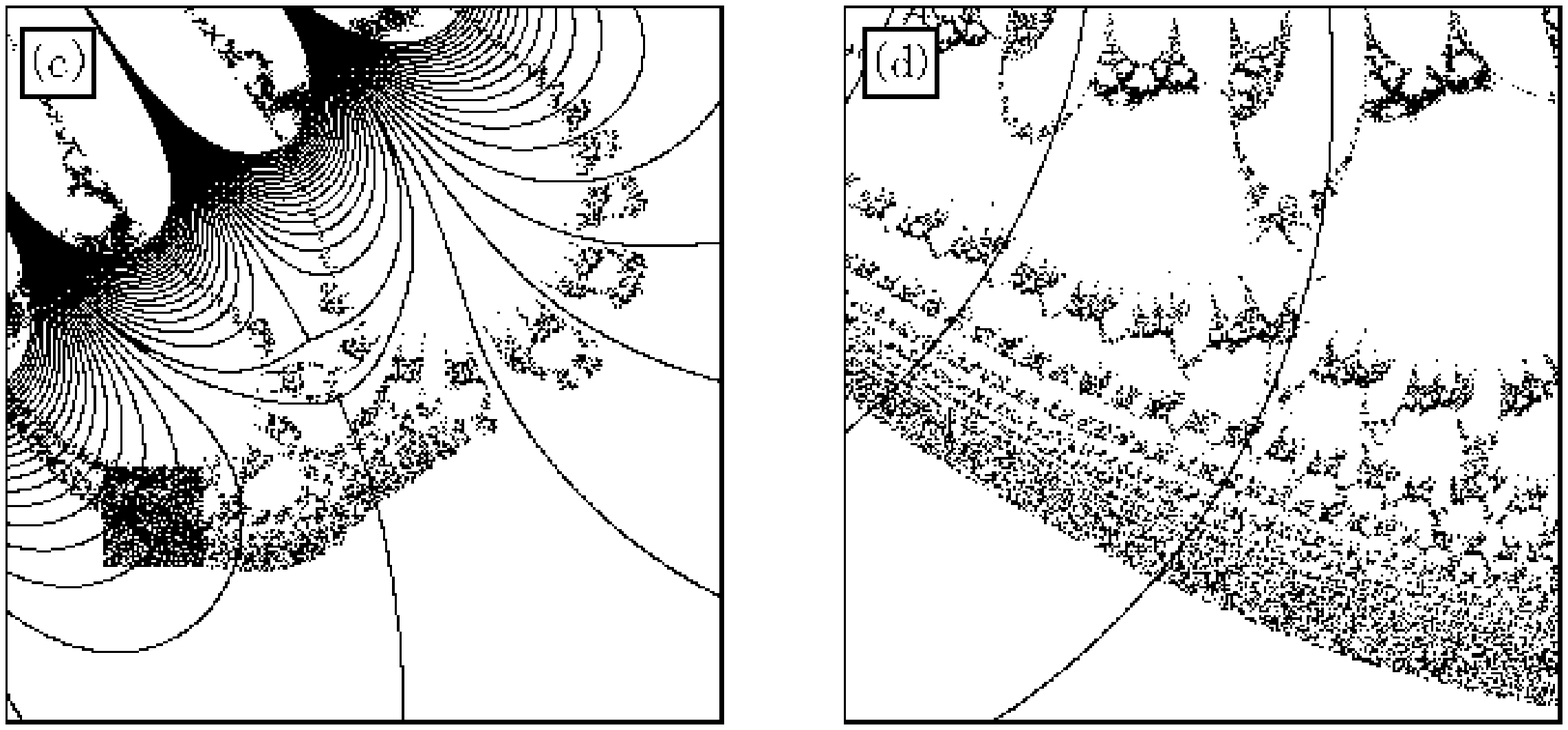}{0.5}{1.0}
\caption{\label{fig_leaves}
Boundaries of (a) the partition ${\cal P}$ and
(b) its forward iteration $ f ({\cal P})$
plotted on ${\cal W}^s(0,0)$, with homoclinic points superposed.
The generations of the homoclinic points displayed are
lower than or equal to 1.
(c) The enlarged figure of the hatched part of (b).
The boundaries of $ f ({\cal P})$
fails to divide the set of homoclinic points clearly,
in the hatched part of (c).
(d) The enlarged figure of the hatched part of (c).
Homoclinic points are aggregated densely like a thick band.
The boundary curves might touch the set of homoclinic points.
}
\end{figure}

Fig.~\ref{fig_leaves} shows the $\xi_s$ plane
divided by $ f ^n({\cal P})$ for $n=0$ and 1.
The central domain in Fig.~\ref{fig_leaves}(a)
is the domain $D$, as is explained in
Sec. \ref{title_emergence_of_homoclinic_tanglement}.
The domain $D$ is defined as a connected domain in
$U(0,0,0)\cap {\cal W}^s(0,0)$
which includes the origin $(0,0)$.
If $ f ^n({\cal P})$ for any $n\ge 0$
divides clearly the set of homoclinic points displayed here,
then ${\cal P}$ is the generating partition, i.e.,
the relation (\ref{def_generating_partition}) holds
for ${\cal P}$ 
(it is sufficient to consider the case of $n\ge 0$
since any homoclinic point is mapped to the region displayed
in Fig.~\ref{fig_leaves}(a)
by the iterations of the map).

On one hand, 
$ f ^n({\cal P})$ for $n=0$
divides a set of homoclinic points clearly
as shown in Fig.~\ref{fig_leaves}(a).
This means that our partition is a reasonable
approximation of the generating partition ${\cal G}$.
So by means of ${\cal P}$,
we can construct a symbolic dynamics which works effectively.
On the other hand, 
there are some regions in the $\xi_s$ plane where
$ f ^n({\cal P})$ for $n\ge 1$ fails to divide
the set of homoclinic points clearly.
In such regions, it may be necessary to improve
our partition to obtain the generating partition,
and it is our future problem. 
Note that our complex dynamics is not proven 
to be sufficiently chaotic, i.e., hyperbolic, 
and that the existence of the generating partition 
for non-hyperbolic systems is an open problem. 
The improvement of the partition is of a mathematical interest,
however,  as actually demonstrated below, the present definition of ${\cal
P}$ is sufficient 
for our semiclassical analysis.

In terms of the partition ${\cal P}$,
each homoclinic point is encoded into a bi-infinite symbolic sequence
of the form:
\begin{eqnarray}\label{def_symbolic_sequence}
& & \!\!\!\!\!\!\!\!\!\!\!\ldots O\,\,O\,a_{-n}\,a_{-(n-1)}\ldots
a_{-1}\,\, .\,\,a_0\,\,\,a_1\ldots a_n\,\,O\,\,O\,\ldots ,
\end{eqnarray}

\noindent
where $O = (0,0,0)$,
$n\in {\mathbb N}$, and $a_k\in {\cal S}\,\,(|k|\le n)$.
The symbol $a_k$ stands for a phase space component denoted by $U(a_k)$
which contains the $k$-th iteration of the homoclinic point,
whose initial location is in $U(a_0)$.
The finite sequence of $a_k\,\,(|k|\le n)$
is accompanied with semi-infinite sequences of $O$'s
in both sides. 
It reflects that any homoclinic point approaches the origin
of phase space by forward and backward iterations of the map
$ f $. 
In particular, 
homoclinic points included in $D$ are encoded into
symbolic sequences of the form:
\begin{eqnarray}
& & \ldots O\,\,O\,\,a_{-n}\,a_{-(n-1)}\ldots
a_{-1}\,\, .\,\,O\,\,O\ldots ,
\end{eqnarray}

\noindent
where $n\in {\mathbb N}$ and $a_k\in {\cal S}\,\,(0<k\le n)$. 
Any symbol in the r.h.s. of the decimal point is $O$, 
since the forward trajectory is always included in a component $U(O)$ 
due to the definition of $D$.

Further investigation of symbolic dynamics,
which is presented in Sec. \ref{title_symbolic_dynamics},
enables us to estimate
the amount of imaginary parts of actions gained by
homoclinic orbits.
Let $w$ a homoclinic point
which has a symbolic sequence of the form (\ref{def_symbolic_sequence}).
When we denote $a_k=(x_k, y_k, \nu_k )$ for $k\in {\mathbb Z}$,
the imaginary part of action gained by the forward trajectory of $w$,
denoted by $s(w)$, is estimated as follows:
\begin{eqnarray}
s(w) &=&
\sum_{k=1}^{+\infty}\hbox{Im}\left[ T(p_{k-1}) - V(q_k)\right]
\nonumber \\
&\approx& \frac{\pi}{\gamma}\sum_{k=1}^{+\infty}
    (x_k\nu_k^{1/2} - x_{k-1}\nu_{k-1}^{1/2})
    (y_k\nu_k^{1/2} - y_{k-1}\nu_{k-1}^{1/2}) ,
\nonumber\\
& & \label{estimation_formula_of_ImS}
\end{eqnarray}

\noindent
where $\gamma$ is a parameter of the potential function $V(q)$.

From these results,
we discuss the symbolic description of
semiclassical candidate orbits.
To this end, we display schematically 
the configurations of homoclinic points 
and the elements of $M$, 
which are shown in the previous 
Figs. \ref{fig_homoclinic_pts}(a) and 
\ref{fig_domain_D_used_in_this_study}(a). 
The configurations can be seen clearly 
by introducing graphs on the $\xi_s$ coordinate 
of ${\cal W}^s(0,0)$, 
whose vertices represent the homoclinic points 
or the elements of $M$.

\begin{figure}
\outputfig{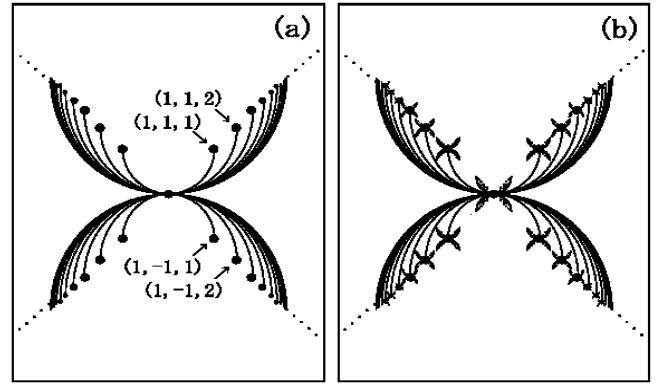}{0.6}{1.0}
\caption{\label{fig_graph1}
Graphs (a) $K_1$ and (b) $K_2$ representing
homoclinic points which have symbolic sequences of forms
$\ldots O\,O\, a_{-1}\,\, .\,\,O\,O\ldots$ and
$\ldots O\,O\, a_{-2}\,\,a_{-1}\,\, .\,\,O\,O\ldots$ respectively,
where $a_{-2},\,a_{-1}\in {\cal S}$.
In (a), the center vertex corresponds to the case of $a_{-1} =O$,
and the vertices in the upper-right side and in the lower-right one
correspond respectively to the cases of $a_{-1} =(1,1,\nu)$ and
$a_{-1} =(1,-1,\nu)$ with $\nu\in {\mathbb N}$, and so on.
}
\end{figure}

The configuration of the homoclinic points
in the domain $D$ is represented by
a graph $K_{\infty}$ defined as follows.
First, by a graph $K_1$ which is shown in Fig.~\ref{fig_graph1}(a),
we represent homoclinic points which have symbolic sequences
of the form:
\begin{eqnarray}
\ldots O\,\,O\,\, a_{-1}\,\, .\,\,O\,\,O\ldots \quad (a_{-1}\in {\cal S}).
\end{eqnarray}

\noindent
Second, we attach smaller copies of $K_1$ to each vertex
of the original $K_1$ to obtain a graph $K_2$ which is shown
in Fig.~\ref{fig_graph1}(b).
By the graph $K_2$, we represent homoclinic points
which have symbolic sequences of the form:
\begin{eqnarray}
\ldots O\,\,O\,\, a_{-2}\,\,a_{-1}\,\, .\,\,O\,\,O\ldots
\quad (a_{-2},a_{-1}\in {\cal S}).
\end{eqnarray}

\begin{figure}
\outputfig{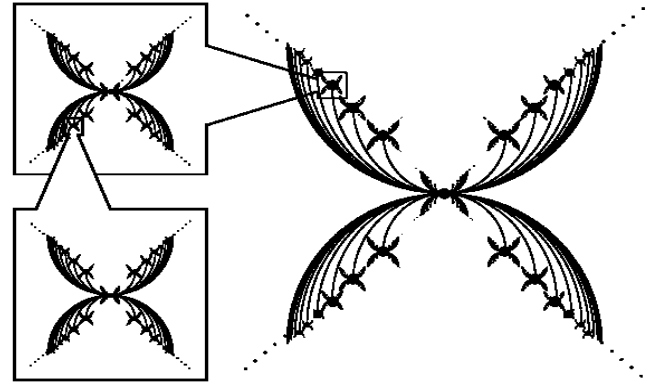}{0.6}{1.0}
\caption{\label{fig_graph3}
The graph $K_{\infty}$ representing homoclinic points
included in the domain $D$.
Hierarchical configurations of vertices
can be seen by enlarging the neighbourhood
of any vertex. 
}
\end{figure}

\noindent
In a similar way, a graph $K_n$ is obtained for any integer $n\ge 3$
by attaching much smaller copies of $K_1$ to each vertex of $K_{n-1}$.
Since inclusions hold as $K_1\subset K_2\subset K_3\subset\ldots$,
we finally obtain a graph $K_{\infty}$ which is shown in Fig.
\ref{fig_graph3}, 
as the union of the graphs $K_1,\,K_2,\,K_3,\ldots$.
$K_{\infty}$ is a tree graph, i.e.,
there is no loop on the graph.

\begin{figure}
\outputfig{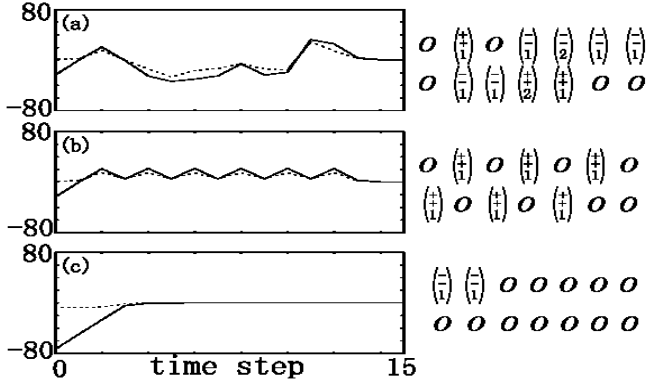}{0.6}{1.0}
\caption{\label{fig_orbits_and_symbols}
The orbits launching from $M$,
and symbolic sequences assigned to the initial points of the orbits.
The upper and lower signs in each symbol denotes $x$ and $y$ 
in (\ref{def_partition_and_symbolic_space}) respectively,
and the integer denotes the value of $\nu$.  
A semi infinite part,
$O\,\,O\,\,O\ldots$, is omitted in each sequence.
Solid and broken lines
represent $\hbox{Re}\,q$ and $\hbox{Im}\,q$ respectively.
The trajectories show (a) erratic motions,
(b) approximately 2-periodic motions, and
(c) monotonical approach to real phase space.
}
\end{figure}

Next we consider the configuration of the elements of $M$.
As shown in Figs. \ref{fig_homoclinic_pts}(a) and 
\ref{fig_domain_D_used_in_this_study}(a),  
there are clear similarities between
configurations of homoclinic points and the elements of $M$.
We define $M_n$ as a subset of $M$
such that the members belong to generations lower than $n$.
Then we observed numerically the following two facts:
one is that on ${\cal W}^s(0,0)$,
the elements of $M_n$ for $n\ge 3$ have the same configuration
as a graph $K_{n-2}$
($M_n$ for $n=2$ has only a single element, and
$M_n$ for $n=1$ is null).
The other is that on the plane ${\cal I}$,
the elements of $M_n$ are located at the centers
of chain-like structures in ${\cal M}_n$
(see Fig.~\ref{fig_M_set}(c)).

In the case of homoclinic points,
each vertex of $K_{n-2}$ corresponds 
to a symbolic sequence of the form:
\begin{eqnarray}
\ldots O\,\,O\,\, a_{-(n-2)}\,\,a_{-(n-3)}\ldots \,a_{-1}
\,\, .\,\,O\,\,O\,\ldots .
\end{eqnarray}

\noindent
Then to each point of $M_n$,
one can assign formally a semi infinite symbolic sequence 
of the form:
\begin{eqnarray}
a_{-(n-2)}\,\,a_{-(n-1)}\,\,\ldots \,\,a_{-1}\,\,O\,\,O\ldots .
\label{sym_seq_assigned_to_I_and_W_no1}
\end{eqnarray}

\noindent
Fig.~\ref{fig_orbits_and_symbols} shows
some trajectories launching from $M$
and symbolic sequences of the form
(\ref{sym_seq_assigned_to_I_and_W_no1})
assigned to the initial points of the trajectories.
The signs and amplitudes of the $q$ components at each time step
are well described by 
$(x,y)$'s and $\nu$'s of 
the corresponding symbols.
This means that there are clear similarities
between motions of both trajectories launching
from $M$ and a set of homoclinic points.
More precisely, 
for an element of $M$ which has a symbolic sequence of the form
(\ref{sym_seq_assigned_to_I_and_W_no1}),
the forward trajectory is well approximated by that
of a homoclinic point which has a symbolic sequence of the form:
\begin{eqnarray}
\ldots O\,\,O\,\, .\,\, a_{-(n-2)}\,\,a_{-(n-1)}\ldots \,a_{-1}
\,\,O\,\,O\,\ldots .
\end{eqnarray}

In this way, the elements of $M_n$ are described by symbolic sequences.
To the chain-like structure in ${\cal M}_n(\subset {\cal I})$
which is associated with each element of $M_n(\subset {\cal W}^s(0,0))$,
we assign the same symbolic sequence as the element of $M_n$,
since as stated in Sec. \ref{title_multi_generation_structure},
the motions of trajectories launching from a single chain-like structure
are well approximated,
till they start to spread along the real domain ${\cal W}^u(0,0)$,
by the motion of a trajectory launching from
an element of $M$ located
at the center of the chain-like structure.
Then chain-like structures in ${\cal M}_n$ are also described
by symbolic sequences of the form (\ref{sym_seq_assigned_to_I_and_W_no1}).

Finally, we discuss the estimation of imaginary parts of actions
gained by semiclassical candidate orbits.
By making use of 
the above mentioned similarities between
$M$ and a set of homoclinic points,
for an element of $M_n(\subset M)$
which has a symbolic sequence:
\begin{eqnarray}
a_0\,\,a_1\,\,a_2\ldots a_{n-3}\,\,O\,\,O\ldots ,
\label{sym_seq_assigned_to_I_and_W_no2}
\end{eqnarray}

\noindent 
(note that the notation is slightly different from
(\ref{sym_seq_assigned_to_I_and_W_no1}))
we evaluate the imaginary part of action of 
its orbit by a homoclinic orbit with a symbolic sequence:
\begin{eqnarray}
\ldots O\,\,O\,\, . \,\,a_0\,\,a_1\,\,a_2\ldots a_{n-3}\,\,O\,\,O\ldots .
\end{eqnarray}

\noindent
Due to this evaluation,
the former imaginary part of action is estimated by
applying (\ref{estimation_formula_of_ImS})
to the symbolic sequence (\ref{sym_seq_assigned_to_I_and_W_no2}).

For each trajectory launching from a single chain-like structure
in ${\cal M}_n$, 
we approximate the imaginary part of action as that of
the trajectory launching from the element of $M_n$
located at the center of the chain-like structure.
In this way, we can estimate the imaginary parts of actions
gained by semiclassical candidate orbits.
Though there are branches in ${\cal M}_n$ which are not included
in any chain-like structure,
semiclassical contributions from them are negligible.
See the discussion in Sec. \ref{title_reproduction_of_tunneling_amplitudes}.

\subsection{Reproduction of Tunneling Wave Functions}
\label{title_reproduction_of_tunneling_amplitudes}

Semiclassical wave functions are constructed
from
significant tunneling orbits
selected according to the amounts of
imaginary parts of actions.
The semiclassical mechanism of the tunneling process
is explained by
the structure of complex phase space.

The wave function 
$\langle q_n | U^n | q_{\alpha}, p_{\alpha} \rangle$ 
given in (\ref{Feynman_integral}) is constructed 
by two steps: 
First, out of $M_n (\subset M)$ the elements are picked up  
such that the imaginary parts of actions
estimated by (\ref{estimation_formula_of_ImS}) are minimal. 
Second, the Van Vleck's formula in (\ref{Van_Vleck_formula_a}) 
is evaluated for the initial points in chain-like structures 
associated with these elements of $M_n$. 
Before picking up the elements out of $M_n$,
the Stokes phenomenon
\cite{Heading} 
should be taken account of.
Following
the prescription given 
in Ref.~\cite{ShudoIkeda3},
we found that chain-like structures
which have unphysical contributions
to the wave function are associated with elements of $M_n$
whose symbolic sequences include symbols
of the form $(1,-1,\nu)$ or $(-1,1,\nu)$ 
with $\nu\in {\mathbb Z}$.
At least, the chain-like structures 
which attain exponentially large semiclassical amplitudes 
are in this case. 
The justification is our future problem.
See, the Discussion.

After excluding such elements from $M_n$,
we have an ordering for elements of $M_n$:
\begin{equation}
w_0\,\,\,\,
w_1\,\,\,\,
w_2\,\,
\ldots ,
\label{ordering_of_symbolic_sequences_a}
\end{equation}     

\noindent
such that inequalities hold:
\begin{equation}
0\le
s(w_0) \le
s(w_1) \le
s(w_2) \le
\ldots ,
\label{ordering_of_symbolic_sequences_b}
\end{equation}

\noindent
where $s(w_k)$ for $k=0,1,2,\ldots$
represents the imaginary part of action
estimated by (\ref{estimation_formula_of_ImS})
for the forward trajectory of $w_k$.

From the Appendix II, 
it is seen that  
$w_0$, which is primarily significant in $M_n$ for the wave function, 
has a symbolic sequence:
\begin{equation}
O\,\,O\,\,O\ldots ,
\label{significant_sequences_primary}
\end{equation}

\noindent
and the members of $\{w_1, w_2, \ldots , w_{2(n-2)} \}$, 
which are secondarily significant in $M_n$, 
have symbolic sequences of the form:
\begin{equation}
b\,\,\,b\,\,\,b\,\ldots \,b\,\,\,b\,\,\,O\,\,O\,\,O\ldots ,
\label{significant_sequences_secondary}
\end{equation}

\noindent
where $b=$ $(1,1,1)$ or $(-1,-1,1)$.
Due to (\ref{sym_seq_assigned_to_I_and_W_no1}),
the length of $\,b\,\,b\,\,b\,\ldots\,b\,\,b\,$ varies from 1 to $n-2$,
and the member of $\{w_1, w_2, \ldots , w_{2(n-2)} \}$
which has $\,b\,\,b\,\,b\,\ldots\,b\,\,b\,$ of length $k-2\,\,(k\le n)$
belongs to the $k$-th generation.
From (\ref{estimation_formula_of_ImS}) 
it is estimated that $s(w_1)=s(w_2)=\ldots =s(w_{2(n-2)})$.

We observed that 
the orbits $\{w_0,\,\,w_1,\ldots ,w_{2(n-2)\}}$
behave as follows. 
The orbit $w_0$  
converges to the origin of phase space exponentially, so that
it gains only small imaginary part of action.
The orbits $\{w_1,\,\,w_2,\ldots ,w_{2(n-2)}\}$
first explore the vicinity of real phase space till
the sequence $\,b\,\,b\,\,b\ldots b\,\,b$ terminates,
and then converge to the origin exponentially.
Such motions yield much smaller imaginary parts of actions
than flipping motions in complex domain.
The latters are generic trajectories launching from $M$.
The motions of homoclinic orbits
whose symbolic sequences include sub sequences
of the form (\ref{significant_sequences_secondary}) are investigated
in Sec. \ref{title_symbolic_dynamics}.
There it is found that such homoclinic orbits
explore the vicinity of real phase space.
The motions of the orbits of $w_1,\,\,w_2 ,\ldots ,w_{2(n-2)}$
reflect those of homoclinic orbits.

\begin{figure}
\outputfig{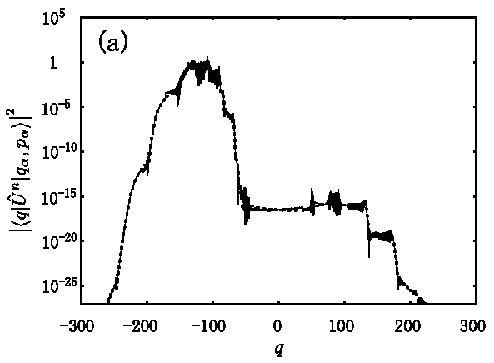}{0.6}{1.0}
\outputfig{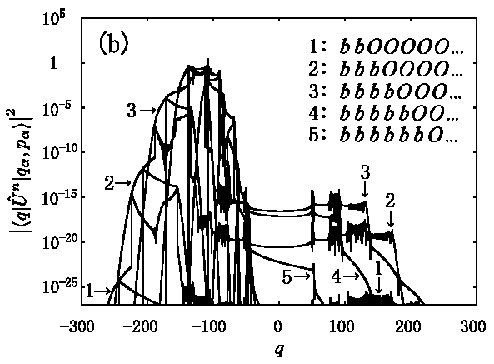}{0.6}{1.0}
\caption{\label{fig_quantum_and_semicls}
(a) 
The squared amplitudes of the
quantum and semiclassical wave functions
for $n = 10$, represented by
dotted and solid curves respectively.
Note that both curves are superposed.
(b)
Individual contributions from chain-like structures
to the semiclassical wave function shown in (a).
For some components, 
symbolic sequences assigned to chain-like structures 
are presented, where $b=(-1,-1,1)$. 
}
\end{figure}

\begin{figure}
\outputfig{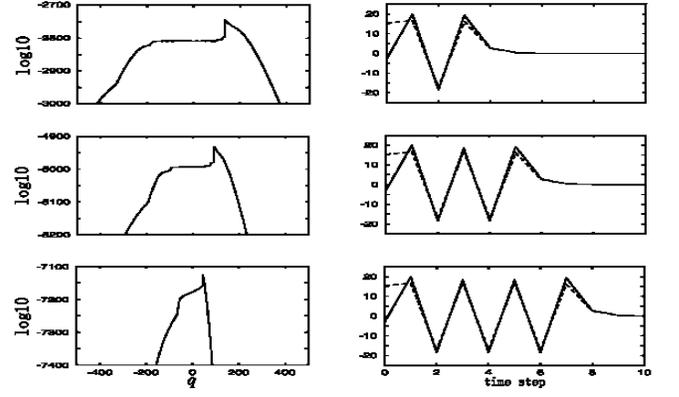}{0.6}{1.0}
\caption{\label{fig_contrib_from_high_gen}
Squared amplitudes of the semiclassical components
(the l.h.s.) which come from chain-like structures
associated with the elements of $M$
whose trajectories exhibit oscillatory motions (the r.h.s.).
Solid and broken lines in the r.h.s. represent $\hbox{Re}\,q$
and $\hbox{Im}\,q$ respectively.
Flipping or oscillatory motions in complex phase space
gain large amounts of
imaginary parts of actions due to large complex momenta.
}
\end{figure}

Fig.~\ref{fig_quantum_and_semicls} shows
quantum and semiclassical wave functions for $n = 10$,
the latter of which is constructed
by taking account of the contributions from
chain-like structures associated with
$w_0,\,\,w_1,\ldots ,w_{2(n-2)}$.
Both functions are in excellent agreement.
The contributions from the other chain-like structures are much smaller
than those from the chain-like structures associated with
$w_0,\,\,w_1,\ldots ,w_{2(n-2)}$.
In fact, the squared amplitudes of them are of order $\sim 10^{-50}$ at
most. 
In particular, the contributions from trajectories which exhibit
flipping or oscillatory motions are negligible, as shown in
Fig.~\ref{fig_contrib_from_high_gen}.

On one hand, 
significant chain-like structures increase
linearly with the time step $n$ as seen above, and 
the branches included in individual chain-like structures 
also increase linearly with $n$, reflecting the oscillations of
real domain ${\cal W}^u(0,0)$. 
So the numbers of branches which we should take into account
increases algebraically with $n$. 
On the other hand, 
the number of chain-like structures in ${\cal M}_n$, and 
that of branches in ${\cal M}_n$ increase exponentially with $n$.
Therefore only a small number of branches are significant
to describe the tunneling process.
The algebraic increase of the significant orbits 
is a consequence of the absence of real-domain chaos. 
When the real domain is chaotic, 
a small piece of manifold which has approached 
the real domain are stretched and folded 
without gaining additional imaginary part of action, 
so that the number of significant orbits 
can increase exponentially with time.

The contributions from branches not included
in any chain-like structures are also negligible,  
such as the branches in 
Fig.~\ref{fig_creation_of_chain_around_intersection}(a) 
except the central five ones.  
For orbits launching from such branches in ${\cal M}_n$,
there is no orbit launching from $M$
which guides them to real phase space within $n$ time steps.
This means that these orbits have large imaginary parts of momenta
at the time step $n$, so that they
gain sufficiently large amounts of
imaginary parts of actions.
If the imaginary parts of actions are positively large,
the contributions from the orbits are small enough to be negligible, or
if they are negatively large,
the contributions from the orbits are unphysical due to the Stokes
phenomenon, 
so that the orbits should be excluded from the whole candidates.

The excellent agreement between both quantum and semiclassical
calculations enables us to interpret semiclassically
the features of tunneling wave functions.
Fig.~\ref{fig_quantum_and_semicls}(b) shows
a decomposition of the semiclassical wavefunction in
Fig.~\ref{fig_quantum_and_semicls}(a)
into components which come from individual chain-like structures.
It can clearly be seen that
the contributions from many chain-like structures reproduce
the crossover of amplitudes in the reflected region and
the staircase seen in the transmitted region.
We found that erratic oscillations on each component
are due to the interferences between branches
included in a single chain-like structure.
The symbolic sequences of the form
(\ref{significant_sequences_secondary})
can have various lengths of $b\,\,b\,\,b\ldots b\,\,b$.
The lengths of this part correspond to the generations of
$w_1,\,\,w_2,\ldots ,w_{2(n-2)}$.
Thus both the staircase and the crossover of amplitudes
are created by the interferences between
chain-like structures belonging to different generations.
In this way, the complicated tunneling amplitude is explained
semiclassically by the creations of chain-like structures
on the initial-value plane,
and by the exponential increase of the number of 
chain-like structures with time 
(though linear increase for significant ones), 
which is due to the emergence of complex homoclinic tanglement.

The semiclassical mechanism of the tunneling process
in our system is summarized as follows.
Stable and unstable manifolds of a real-domain unstable orbit
create tanglement in complex domain.
The initial manifold representing a quantum initial state is
located in the tanglement, so that the intersection points between
the initial manifold and the stable manifold form
a hierarchical structure
on the initial manifold.
The orbits launching from the neighbourhood of each intersection point
are guided to real phase space by the stable manifold, and then
spread over the unstable manifold.
The number of the orbits guided to real phase space 
increases exponentially with time
(though significant ones increase algebraically),
reflecting the hierarchical structure formed by
the intersection points on the initial manifold.
Then the interferences between these orbits create
complicated patterns in the tunneling amplitude.

\section{Symbolic Description of Complex Homoclinic Structure}
\label{title_symbolic_dynamics}

\subsection{Construction of Partition of Phase Space}
\label{title_construction_of_partition}

Here we construct a partition of phase space
which encodes homoclinic points 
of the origin $(0,0)$ 
into symbolic sequences
and defines a symbolic dynamics which works effectively.
The behaviors of homoclinic orbits and 
the estimation of imaginary parts of actions for the orbits 
are presented in the following Secs. 
\ref{title_properties_of_homoclinic_orbits} and 
\ref{title_estimation_of_imaginary_actions} respectively.

There has been extensive studies on
the construction of generating partition
in real phase space
\cite{Cvitanovic}
even in non-hyperbolic regimes
\cite{ChristiansenPoliti}.
In such real-domain studies,
the boundaries of generating partition
are roughly 
approximated 
by a set of folding points created by
the single-step iterations of flat manifolds.
However, the extension of such working principle
to complex phase space is not obvious.

\begin{figure}
\outputfig{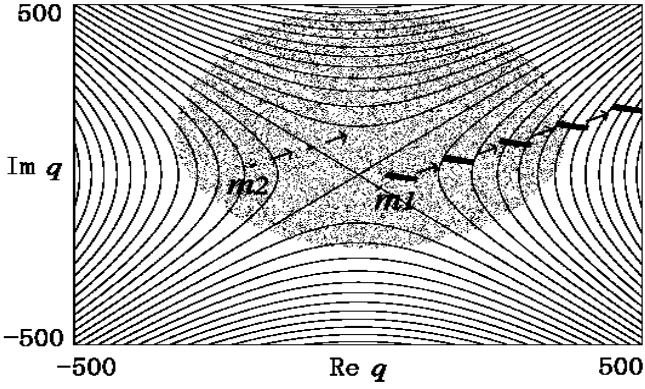}{0.6}{1.0}
\caption{\label{fig_asym_bhv}
Iterations of small pieces of a flat manifold given by
$\{(q,p)\in {\mathbb C}^2\,\,|\,\,p=p_0\}$ 
for $p_0=100.0+{\rm i}\,50.0$,
with contour curves of the $u$ component of the $(u,v)$ coordinate
superposed.  
One of the small pieces, $m_1$, is hardly deformed by the iterations of
the map, and remains to be at almost the same momentum as
initially given. 
The other one, $m_2$, expands over a wide range of phase space 
(a hatched region) after a few step iterations.
}
\end{figure}

\begin{figure}
\outputfig{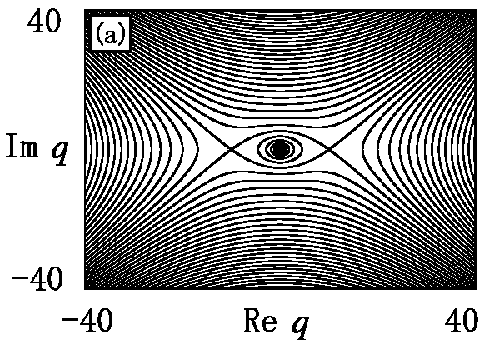}{0.6}{1.0}
\outputfig{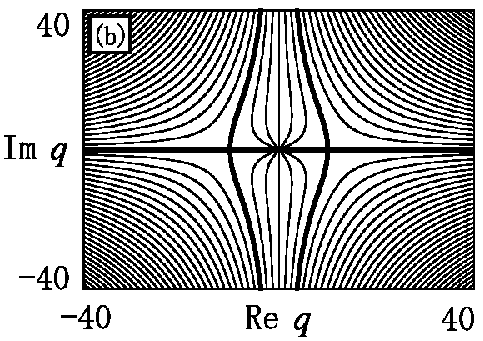}{0.6}{1.0}
\caption{\label{fig_contour_maps}
Contour curves of the functions (a) $A(q)$ and (b) $B(q)$.
In (a), each curve is given by
$A(q)=\left\{1+\log (2\gamma)\right\}/2 +0.4 n$ for
$n\in {\mathbb Z}$,
which approaches the axes
$\hbox{Re}\,q=\pm\hbox{Im}\,q$
in both directions of the curve.
In (b), each curve is given by
$B(q)= \pi n/6$ for $n\in {\mathbb Z}$
with the branch $-\pi\le\hbox{Arg}\,q<\pi$.
Bold curves 
in the side of $\hbox{Re}\,q >0$ correspond to the case of
$B(q)=0$, and 
those in the other side corresponds to the cases of
$B(q) = -\pi\,(\hbox{Im}\,q\ge 0)$ and $\pi\,(\hbox{Im}\,q\le 0)$.
The bold curves 
intersect the real axis at $q=\pm 1/\sqrt{2\gamma}$.
}
\end{figure}

In order to know the locations of the boundaries
of the generating partition,
we consider the iterations of flat manifolds given by the form
$\{(q,p)\in {\mathbb C}^2\,|\,p=p_0\}$ for a complex
value $p_0$. 
Fig.~\ref{fig_asym_bhv} shows the iterations
of some small pieces of
such a flat manifold, which exhibit a variety of behaviors
depending on the initial locations of the small pieces.
Since the magnitude and phase of the gradient $V^{\prime}(q)$
are controlled respectively by the functions $A(q)$ and $B(q)$
which appear in (\ref{phase_of_gradient}),
the differences of the behaviors shown in the figure mainly come from
the differences of the values of $A(q)$.
The contour curves of $A(q)$ and $B(q)$ are shown in
Fig.~\ref{fig_contour_maps}.

In order to know the locations of the boundaries of the partition
far from the origin $(0,0)$,
we introduce a coordinate $(u,v)$ in the $q$ plane:
\begin{subequations}\label{definition_of_hyp_cdn}
\begin{eqnarray}
u & = & \left[ (\hbox{Re}\,q)^2 -(\hbox{Im}\,q)^2\right] /2 ,
\label{definition_of_hyp_cdn_a}\\
v & = & \hbox{Re}\,q\cdot \hbox{Im}\,q .
\label{definition_of_hyp_cdn_b}
\end{eqnarray}
\end{subequations}

\noindent
On this coordinate,
we obtain the estimations:
\begin{subequations}\label{estimations_of_amplitude_and_phase}
\begin{eqnarray}
A(q) &=& \gamma \left[ (\hbox{Re}\,q)^2 -(\hbox{Im}\,q)^2\right] 
         -\log |q| \nonumber \\
     &=& 2\gamma u + O\left(\log |u|\right) \quad\!\!\!\!\!
         \left( v:\hbox{fixed},\,\, |u|\,\rightarrow\,\infty \right)\!, 
         \ \ \ \ \label{amplitude_estimation} \\
B(q) &=& 2\gamma \hbox{Re}\,q\cdot \hbox{Im}\,q 
         - \hbox{Arg}\,\,q\nonumber \\
     &=& 2\gamma v + O\left(\log |v|\right) \quad\!\!\!\!\!
         \left( u:\hbox{fixed},\,\, |v|\,\rightarrow\,\infty \right)\!, 
         \ \ \ \ \label{phase_estimation}\\
\hbox{or}
     & & 2\gamma v + O\left(|u|^{-1}\right) \quad\!\!\!
         \left( v:\hbox{fixed},\,\, |u|\,\rightarrow\,\infty \right)\!, 
         \ \ \ \ \label{phase_estimation2}
\end{eqnarray}
\end{subequations}

\noindent
where $\gamma$ is a parameter of $V(q)$.
The above estimations imply that far from the origin $(0,0)$
the magnitude and phase of $V^{\prime}(q)$
are controlled by $u$ and $v$ respectively.

On one hand, 
in a region of phase space with $u\gg 1$, 
which includes the real asymptotic region, 
$V^\prime (q)$ almost vanishes so that 
the behavior of a small piece of manifold there, 
typically observed as that of $m_1$ in Fig.~\ref{fig_asym_bhv}, 
is of a free motion. 
On the other hand, 
in a region of phase space with $u\ll -1$,  
any small rectangle on the $(u,v)$ coordinate,  
$\{q\in {\mathbb C}\,\,|\,\,
u\in [u_0, u_0+\Delta u],\,v\in [v_0,v_0+\Delta v]\}$, 
with $u_0\ll -1$ is mapped by $V^\prime (q)$ 
approximately in an annulus of radii 
$2\gamma k \exp (-2\gamma u_0)$ and 
$2\gamma k\exp\left[ -2\gamma (u_0+\Delta u)\right]$ 
on the $q$ plane. 
In Fig.~\ref{fig_asym_bhv},
a small piece of manifold $m_2$  
is set as $\Delta v = \pi /\gamma$, 
which approximately corresponds to 
a single period of the phase $B(q)$, so that 
its final manifold looks like an annulus on the $q$ plane
(one radius is much smaller than the other).
Thus when we put the $v$ component of $m_2$
as $[v_0, v_0 + n\pi /\gamma ]$ for an arbitrary integer $n$,
the final manifold of $m_2$ covers 
almost the same range of the $q$ plane
$n$ times, i.e., looks like a superposition of $n$ 
annuli 
on the $q$ plane.
The simplest way to distinguish each 
annulus 
would be to locate a boundary of partition
along the contour curve $v = v_0 + n\pi /\gamma$ on the $q$ plane
for any integer $n$.
From
this observation,
one can expect that far from the origin $(0,0)$,
the boundaries of the partition
should take the form
$\{(q,p)\in {\mathbb C}^2\,|\,v = v_0 + n\pi /\gamma \}$
with $v_0$ and $n$ being some real value and 
arbitrary integer respectively.

\begin{figure}
\outputfig{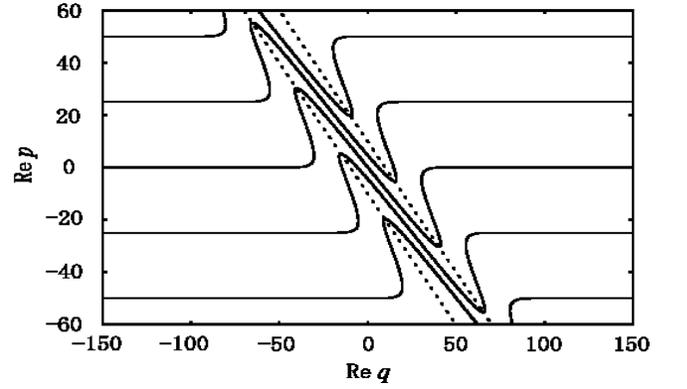}{0.6}{1.0}
\caption{\label{fig_folded_manifolds}
Single-step backward iterations of flat manifolds
given by $\{(q,p)\in {\mathbb C}^2\,\,|\,\,p=p_0\}$
ranging from $p_0=-75.0$ to 75.0 by every 25.0.
The dotted lines represent a set of points
each of which gives the maximal or minimal momentum on each folded manifold.
}
\end{figure}

Around the origin $(0,0)$, 
as shown in Fig.~\ref{fig_folded_manifolds},
the locations of the boundaries 
can be estimated by the single-step backward iterations
of flat manifolds evaluated in real phase space.
The dotted lines represent a set of folding points of the folded manifolds,
each point of which is defined by the condition $dp(q)/dq = 0$
where the differentiation is along each folded manifold.
The set of such folding points are given by
the single-step backward iterations of the lines
$\{(q,p)\in {\mathbb R}^2\,|\,q = \pm 1/\sqrt{2\gamma}\}$,
so that the boundaries of the partition in complex domain are expected
to intersect the real domain around these lines.

We propose a partition of phase space
which satisfies the rough estimations presented above
both in complex and real domains.
To define the partition, we prepare the notations:
\begin{subequations}
\begin{eqnarray}
U(x,y) &=& \{ (q,p)\in {\mathbb C}^2\,
|\, x\,\hbox{Re}\,q>0,\,\, y\,\hbox{Im}\,q>0 \}, \ \ \ \ \\
\beta (x,y,\nu ) &=& \left[ 2\nu xy - (3x+1)y/2\right] \pi ,
\end{eqnarray}
\end{subequations}

\noindent
where $(x,y,\nu)$ is an element of 
${\cal T}\otimes {\mathbb N}$
which appeared in (\ref{def_of_symbols}).
$U(x,y)$ covers a single quadrant of the $q$ plane and
$\beta (x,y,\nu )$ always takes an integer times $\pi$.
For $(x,y,\nu)\in {\cal T}\otimes {\mathbb N}$,
a phase space component $U(x,y,\nu )$ is defined by
\begin{eqnarray}
U(x,y,\nu ) & = & \{  (q,p)\,\, |\,\, (q,p)\in U(x,y), \nonumber\\
& & 
\left[ B(q)-\beta (x,y,\nu)   \right] 
\left[ B(q)-\beta (x,y,\nu +1)\right] \le 0 \} ,
\nonumber\\
& & \label{def_of_phase_space_component_a}
\end{eqnarray}

\noindent
and for $(x,y,\nu)=(0,0,0)$, by
\begin{eqnarray}
U(x,y,\nu) & = & {\mathbb C}^2 -
\bigcup_{(x^{\prime},y^{\prime},\nu^{\prime} )\in
{\cal T}\otimes\hbox{\scriptsize\bf N}}
U(x^{\prime},y^{\prime},\nu^{\prime}).
\label{def_of_phase_space_component_b}
\end{eqnarray}

Then our partition ${\cal P}$ is defined as a set of
the above phase space components, as already shown in
(\ref{def_of_partition}), and
the boundaries of ${\cal P}$ take the form in
(\ref{def_of_boundary_in_Sec_II}).
Such definition of partition satisfies our rough estimation
for the locations of the boundaries.
In fact, 
in the complex domain far from the origin $(0,0)$,
due to (\ref{estimations_of_amplitude_and_phase}),
the relation $B(q) = \beta (x,y,\nu)$ leads to
$v \approx v_0 + n\pi /\gamma$
when we set $v_0 = -(3x+1)y\pi/4\gamma$ and $n=\nu xy$.
Moreover, Fig.~\ref{fig_contour_maps}(b)
shows that the boundaries of ${\cal P}$ indicated by
$B(q)=\beta (x,y,\nu)$ for
$(x,y,\nu)=(1,1,1),(1,-1,1),(-1,1,1,)$ and $(-1,-1,1)$
intersect the real phase space
at $q = \pm 1/\sqrt{2\gamma}$.

\subsection{Properties of Homoclinic Orbits}
\label{title_properties_of_homoclinic_orbits}

By the partition constructed above,
each homoclinic points is encoded into a  sequence of
symbols of the form $(x,y,\nu)$.
In order to estimate imaginary parts of actions
gained by homoclinic orbits,
it is necessary to understand typical behaviors
exhibited by the homoclinic orbits.
Here we present such typical behaviors
as two observations obtained from numerical computations.
The first observation is concerned with the relation between
the integer $\nu$, 
which is a member of a symbol $(x,y,\nu)$
in a symbolic sequence,
and the flipping amplitude of the corresponding trajectory.
The other of them is concerned with the relation between
the length of a consecutive part $b\,\,b\ldots b 
\,\,(b\in {\cal S})$
in a symbolic sequence and the behavior of the corresponding trajectory.
These are numerical observations and
we have no mathematical proof, but
the phase-space itinerary of any homoclinic orbit can
be well explained by the combinations of the behaviors
presented in these observations.

Before presenting the observations, we
see the single-step folding process of flat manifolds
to estimate the locations of homoclinic points in phase space.
Let $m_i$ and $m_f$ be two-dimensional flat manifolds defined by
the conditions $p=p_i$ and $p=p_f$ respectively,
with $p_i$ and $p_f$ being complex
values. 
The intersection $ f (m_i)\cap m_f$
is given by 
$\{ (q,p_f)\in {\mathbb C}^2\,|\, A(q)
=-\log |c|,\,\,B(q)=-\arg c\pm 2\nu\pi,\,\,\nu =0,1,2,\ldots\}$
where $A(q)$ and $B(q)$ are the functions given in
(\ref{estimations_of_amplitude_and_phase}),
and $c = (p_f-p_i)/(2\gamma k)$.
Since the $q$ components of the intersection points on $m_f$ are
located on a single contour curve of $A(q)$,
they are located along the axes $\hbox{Re}\,q=\pm\hbox{Im}\,q$
asymptotically as $\nu\rightarrow +\infty$.
Also the $q$ components of homoclinic points are located along these axes
asymptotically as $\nu\rightarrow +\infty$, where
$\nu$ is a member of a symbol $(x,y,\nu)$ in a symbolic sequence. \\

\noindent
Observation 1. 
Let $\{ w_1, w_2, w_3, \ldots \}$ be a set of homoclinic points
whose symbolic sequences take the forms:
\begin{eqnarray}\label{conditions_of_symbols_in_obs2}
& &\ldots a_{-2}\,\,a_{-1}\,\, .\,\,b_1\,\,\,a_1\,\,\,a_2 \ldots\quad
\hbox{for}\,\, w_1, \nonumber \\
& &\ldots a_{-2}\,\,a_{-1}\,\, .\,\,b_2\,\,\,a_1\,\,\,a_2 \ldots\quad
\hbox{for}\,\, w_2, \nonumber \\
& &\ldots a_{-2}\,\,a_{-1}\,\, .\,\,b_3\,\,\,a_1\,\,\,a_2 \ldots\quad
\hbox{for}\,\, w_3, \nonumber \\
& & \qquad\qquad\qquad \vdots
\end{eqnarray}

\noindent
where $a_k\,(k\ne 0)$ is any member of ${\cal S}$
defined in (\ref{def_partition_and_symbolic_space}),
and $b_{\nu} = (x,y,\nu)$ for $\nu=1,2,3,\ldots$
with $(x,y)\,(\in {\cal T})$ being fixed.
Then the following relations hold:
\begin{subequations}\label{estimation_of_q_and_p_in_obs1}
\begin{eqnarray}
\lim_{\nu\rightarrow +\infty}
\sqrt{\gamma /(\nu\pi )}\,\,q_0(w_{\nu})
 \quad &=& \quad (x,y),
\label{estimation_of_qn_in_obs1}\\
\lim_{\nu\rightarrow +\infty}
\sqrt{\gamma /(\nu\pi )}\,\,p_0(w_{\nu})
 \quad &=& \,\, -(x,y),
\label{estimation_of_pn_in_obs1}\\
\lim_{\nu\rightarrow +\infty}
\sqrt{\gamma /(\nu\pi )}\,\,p_{-1}(w_{\nu})
 \,\, &=& \quad  (x,y),
\label{estimation_of_pn-1_in_obs1}
\end{eqnarray}
\end{subequations}

\noindent
where 
$(q_0,\,p_0)$ is the current location of $w_\nu$ in phase space,
and $p_{-1}$ is the momentum at the last time step.
The r.h.s. of each equation denotes a pair of signs of
real and imaginary parts.\\

\begin{figure}
\outputfig{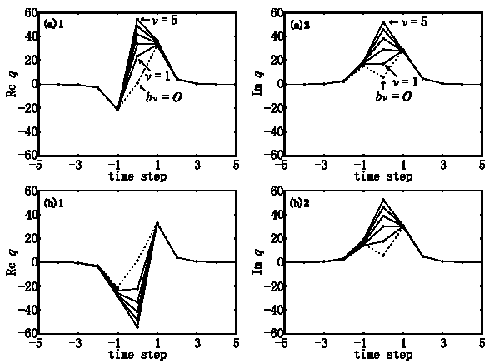}{0.6}{1.0}
\outputfig{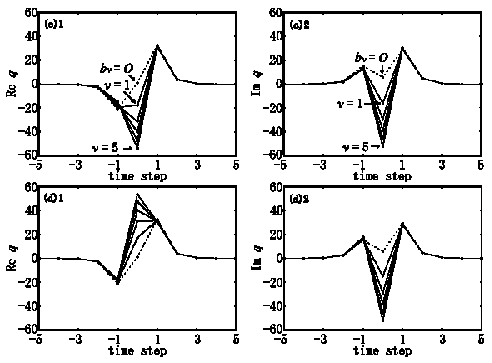}{0.6}{1.0}
\caption{\label{fig_obs1_no1}
The trajectories of homoclinic points $w_{\nu}$
for $\nu=1,2,\ldots ,5$.
The left and right columns display
the real and imaginary parts of the trajectories respectively.
The symbolic sequence of $w_{\nu}$ is given by
$\ldots O\,O\,(-1,1,1)\, .\,\,b_{\nu}\,\,(1,1,2)\,O\,O\,O\ldots$ with
$b_{\nu}$ being 
(a) $(1,1,\nu)$, 
(b) $(-1,1,\nu)$, 
(c) $(-1,-1,\nu)$, and
(d) $(1,-1,\nu)$.  
The dotted lines represent the case that $b_{\nu}=O$.
}
\end{figure}

\begin{figure}
\outputfig{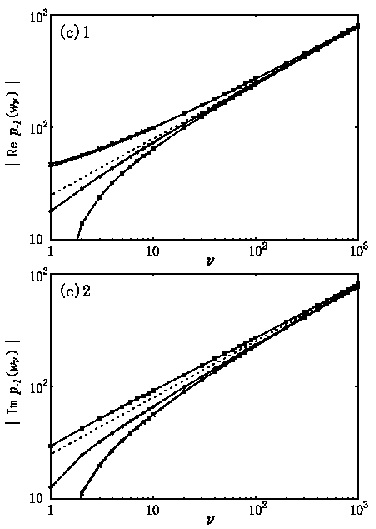}{1.14}{0.95}
\caption{\label{fig_obs1_no2}
The dependences of 
(a) $q_0(w_\nu )$, (b) $p_0(w_\nu )$, and (c) $p_{-1}(w_\nu )$
on the subscript number $\nu$.
The symbolic sequences of $w_{\nu}$ are given in (a),
where $A_{\nu},B_{\nu},$ and $C_{\nu}$ denote
$(1,1,\nu),(-1,1,\nu)$, and $(-1,-1,\nu)$ respectively.
The dotted line in each figure represents
$\sqrt{\nu\pi /\gamma}$.
In (a), all curves almost coincide.
}
\end{figure}

\begin{figure}
\outputfig{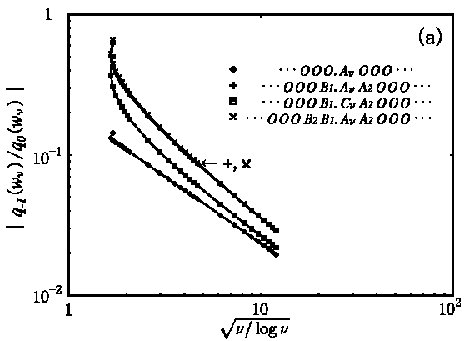}{0.6}{1.0}
\outputfig{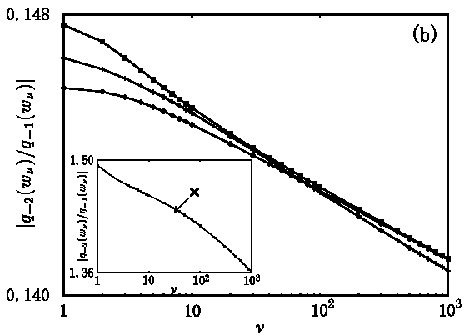}{0.6}{1.0}
\caption{\label{fig_obs1_no3}
The $\nu$ dependence of 
(a) $|q_{-1}(w_{\nu})/q_0(w_{\nu})|$ and
(b) $|q_{-2}(w_{\nu})/q_{-1}(w_{\nu})|$
(the vertical axes of (b) and the inset have 
the common logarithmic scale). 
The symbolic sequences of $w_{\nu}$ are the same as in 
Fig. \ref{fig_obs1_no2}.
Solid lines are the guides for the eye.
}
\end{figure}

Fig.~\ref{fig_obs1_no1} shows the trajectories of $w_{\nu}$'s
for $\nu =1,2,\ldots ,5$.
In these figures, one can see two facts:
one is that 
the signs of $\hbox{Re}\,q_0(w_{\nu})$ and $\hbox{Im}\,q_0(w_{\nu})$
are described respectively by $x$ and $y$ in the symbol $b_{\nu}$.
The other is that 
the amplitudes of $\hbox{Re}\,q_0(w_{\nu})$ and $\hbox{Im}\,q_0(w_{\nu})$
increase with $\nu$ much faster than the amplitudes at the other time steps.
Due to the second fact, 
the following approximations hold for large $\nu$'s:
\begin{subequations}
\begin{eqnarray}
p_0\,\,\,\,    &=& q_1 - q_0\,\,\,\, \approx -q_0 ,\\
p_{-1} &=& q_0 - q_{-1} \approx \,\,\,\,q_0,
\end{eqnarray}
\end{subequations}
so that 
the sign of
$\hbox{Re}\,p_0(w_{\nu})$ (resp. $\hbox{Im}\,p_0(w_{\nu})$)
is opposite to that of
$\hbox{Re}\,p_{-1}(w_{\nu})$ (resp. $\hbox{Im}\,p_{-1}(w_{\nu})$)
for large $\nu$'s. 
Fig.~\ref{fig_obs1_no2} shows
$q_0(w_{\nu})$, $\,p_0(w_{\nu})$, and $p_{-1}(w_{\nu})$
for much larger $\nu$'s.
The magnitudes of the real and imaginary parts of these quantities
are shown to have the dependence of the form $\sqrt{\nu\pi /\gamma}$
for sufficiently large $\nu$'s.
Fig.~\ref{fig_obs1_no3} shows
that $q_0(w_j)$ diverges much faster than $q_{-1}(w_j)$
and $q_{-2}(w_j)$. 
Similarly , $q_0(w_j)$ diverges much faster than $q_k(w_j)$
for any other $k\ne 0$, though that is not displayed here.

\begin{figure}
\outputfig{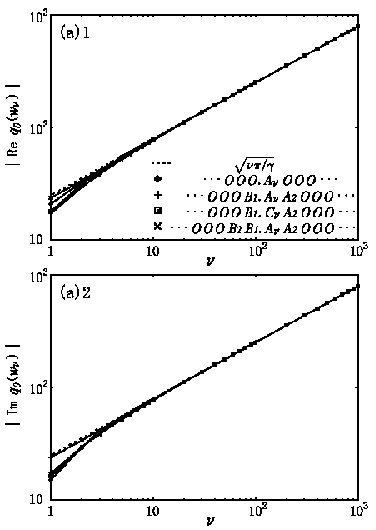}{1.14}{0.95}
\outputfig{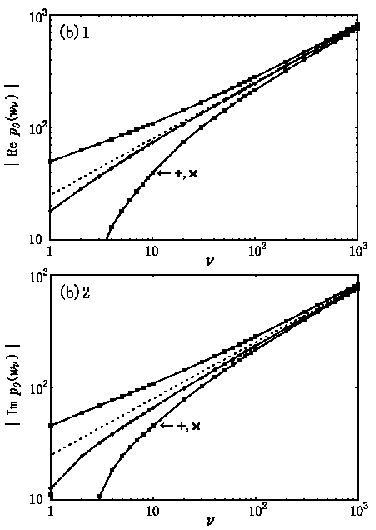}{1.14}{0.95}
\end{figure}

The relation (\ref{estimation_of_qn_in_obs1}) leads to
\begin{eqnarray}\label{asm_relation_in_obs1}
\lim_{\nu\rightarrow
+\infty}|\hbox{Re}\,q_0(w_{\nu})/\hbox{Im}\,q_0(w_{\nu})|=1.
\end{eqnarray}
 
\noindent
In the following, we explain why the relations
(\ref{estimation_of_q_and_p_in_obs1}) should follow,
assuming that the relation (\ref{asm_relation_in_obs1}) holds
for the homoclinic points $w_{\nu}\,(\nu=1,2,3,\ldots)$
given by (\ref{conditions_of_symbols_in_obs2}),
and that $q_0(w_{\nu})$ diverges much faster than
$q_k(w_{\nu})$ for any $k\ne 0$ as $\nu\rightarrow +\infty$.

The relation (\ref{estimation_of_qn_in_obs1}) is explained as follows.
Since $b_{\nu} = (x,y,\nu)$,
$w_{\nu}$ is included in a phase space component $U(x,y,\nu)$
defined in 
(\ref{def_of_phase_space_component_a}) and 
(\ref{def_of_phase_space_component_b}).
Then the $v$ component of $q_0(w_{\nu})$ in the $(u,v)$ coordinate diverges
as $\nu\rightarrow +\infty$, since
$B(q_0(w_{\nu}))$ diverges as $\nu\rightarrow +\infty$
due to (\ref{def_of_boundary_in_Sec_II}), and
$v \approx B(q_0(w_{\nu}))/(2\gamma )\approx 2\nu xy\pi /(2\gamma )$
for large $\nu$'s 
due to (\ref{estimations_of_amplitude_and_phase})
and (\ref{def_of_boundary_in_Sec_II}).
Therefore from (\ref{asm_relation_in_obs1}) and
the relation $\hbox{Re}\,q_0\cdot\hbox{Im}\,q_0=v\approx \nu xy\pi /\gamma $,
we obtain $q_0(w_{\nu})\approx \sqrt{\nu\pi/\gamma}\,(x,y)$
for large $\nu$'s.

The relation (\ref{estimation_of_pn_in_obs1}) is explained as follows.
The classical equations of motions in (\ref{classical_equations})
lead to the relation:
\begin{eqnarray}
q_0(w_{\nu})+q_{-2}(w_{\nu})&=&2q_{-1}(w_{\nu})-V^{\prime}(q_{-1}(w_{\nu})).
\end{eqnarray}
Since $q_0(w_{\nu})$ diverges much faster than $q_{-2}(w_{\nu})$
as $\nu\rightarrow +\infty$ due to our assumption,
the r.h.s. of the above relation diverges in this limit.
It means that $q_{-1}(w_{\nu})$ also diverges
as $\nu\rightarrow +\infty$,
since the r.h.s. of the relation is an entire function of $q_{-1}$.
In particular, 
the $u$ component of $q_{-1}(w_{\nu})$ diverges as $\nu\rightarrow +\infty$,
since $(q_{-1}(w_{\nu}),p_{-1}(w_{\nu}))$ is always included in
the phase-space component $U(a_{-1})$ irrespective of $\nu$.
Furthermore,
the $u$ component diverges to $-\infty$, since if
the $u$ component diverges to $+\infty$
with $(q_{-1}(w_{\nu}),p_{-1}(w_{\nu}))$
being in the same phase-space component, then
$q_{-1}(w_{\nu})$ approaches the real axis and 
$V^{\prime}(q_{-1}(w_{\nu}))$ vanishes, so that 
$\hbox{Im}\left[ q_0(w_{\nu})+q_{-2}(w_{\nu})\right] =
\hbox{Im} \left[ 2q_{-1}(w_{\nu})-V^{\prime}(q_{-1}(w_{\nu}))\right]
\rightarrow 0$.
However, the above contradicts that 
as $\nu\rightarrow +\infty$,
$\left|\hbox{Im}\left[ q_0(w_{\nu})+q_{-2}(w_{\nu})\right]\right|
\approx |\hbox{Im}\,q_0(w_{\nu})|
\approx\sqrt{\nu\pi/\gamma}\rightarrow +\infty$.
Thus the $u$ component of $q_{-1}(w_{\nu})$ diverges to $-\infty$ 
as $\nu\rightarrow +\infty$.
When the $u$ component of $q_{-1}(w_{\nu})$ is negatively large,
$V^{\prime}(q_{-1}(w_\nu))$ is exponentially larger than $q_{-1}(w_{\nu})$,
so that $q_0(w_\nu)\approx -V^{\prime}(q_{-1}(w_\nu))$.
This relation means that $|q_{-1}|\approx \sqrt{{\gamma}^{-1}\log |q_0|}$,
and thus $p_0(w_\nu)$ has the same dependence as $q_0(w_\nu)$ on $\nu$
due to the relation $p_0=q_0-q_{-1}$.
The relation (\ref{estimation_of_pn-1_in_obs1})
is explained in a similar way.

We proceed to the other observation.
In usual symbolic dynamics, a consecutive part $b\,\,b\ldots b$
of a single symbol $b$
in a symbolic sequence always corresponds to a fixed point in phase space
or the motion approaching the fixed point.
However, our classical dynamics always has only a single fixed point $(0,0)$
for any choice of positive parameters $k$ and $\gamma$,
which is easily checked by solving $ f (q,p)=(q,p)$,
so that the phase-space motion corresponding to
a consecutive part $b\,\,b\ldots b$ with $b\ne (0,0,0)$ is not obvious.
Our second observation says that the phase-space motion corresponding to
the above consecutive part 
has a turning point. 
It is conjectured that as the length of $b\,\,b\ldots b$ increases, 
the location of the turning point diverges, so that
the trajectory corresponding to $b\,\,b\ldots b$ 
does not approach to any point in phase space 
in the limit of the length. 
\\

\noindent
Observation 2. 
Let $\left\{ \,w_1,w_2,w_3\ldots\right\}$
be a set of homoclinic points such that
the symbolic sequences take the forms:
\begin{eqnarray}\label{symbolic_sequence_in_property3}
& & \ldots a_{-2}\,\,a_{-1}\,\,.\,\,
    b\,\,\,\,\,b\,\,\,\, a_0\,\,a_1\,\,a_2\ldots\qquad\qquad\qquad\,\,
    \hbox{for}\,\,w_1, \nonumber \\
& & \ldots a_{-2}\,\,a_{-1}\,\,.\,\,
    b\,\,\,\,\,b\,\,\,\,\, b\,\,\,\,\,b\,\,\,\,\,
    a_0\,\,a_1\,\,a_2\ldots\qquad\quad\,\,\,\,
    \hbox{for}\,\,w_2, \nonumber \\
& & \ldots a_{-2}\,\,a_{-1}\,\,.\,\,
    b\,\,\,\,\,b\,\,\,\,\,b\,\,\,\,\,b\,\,\,\,\,\,b\,\,\,\,\,b\,\,\,\,\,
    a_0\,\,a_1\,\,a_2\ldots\quad
    \hbox{for}\,\,w_3, \nonumber \\
& & \qquad\qquad\qquad \vdots
\end{eqnarray}
 
\noindent
where $b\ne (0,0,0)$ and
the length of $b\,\,b\ldots b$ for $w_j\,(j=1,2,3,\ldots )$ is $2j$.
Then the trajectory of $w_j$ corresponding to the consecutive part
$b\,\,b\ldots b$ is included in a single phase-space component $U(b)$,
and the momentum almost vanishes at time step $j-1$.
Moreover, the following inequalities hold:
\begin{subequations}\label{turning_pts_in_property3}
\begin{eqnarray}
& &0< \hbox{Re}\,q_{k-1}(w_j)/\hbox{Re}\,q_k(w_j) < 1 \quad (0< k< \,\,\,j),
\ \ \ \ \ \ \label{turning_pts_in_property3_a}\\
& &0< \hbox{Im}\,q_{k}(w_j)/\hbox{Im}\,q_{k-1}(w_j) < 1 \quad (0< k< \,\,\,j),
\ \ \ \ \ \ \label{turning_pts_in_property3_b}\\
& &0< \hbox{Re}\,q_k(w_j)/\hbox{Re}\,q_{k-1}(w_j) < 1 \quad (j< k< 2j),
\ \ \ \ \ \ \label{turning_pts_in_property3_c}\\
& &0< \hbox{Im}\,q_{k-1}(w_j)/\hbox{Im}\,q_k(w_j) < 1 \quad (j< k< 2j).
\ \ \ \ \ \ \label{turning_pts_in_property3_d}
\end{eqnarray}
\end{subequations}\\

\begin{figure}
\outputfig{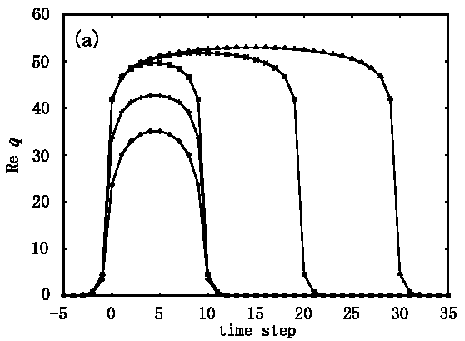}{0.585}{0.975}
\outputfig{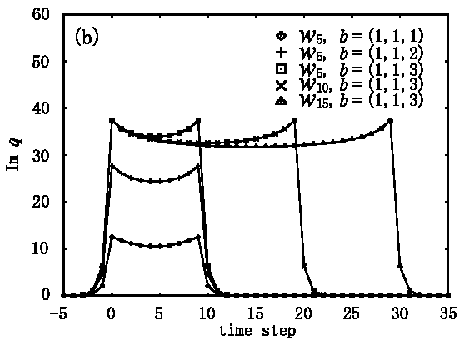}{0.585}{0.975}
\outputfig{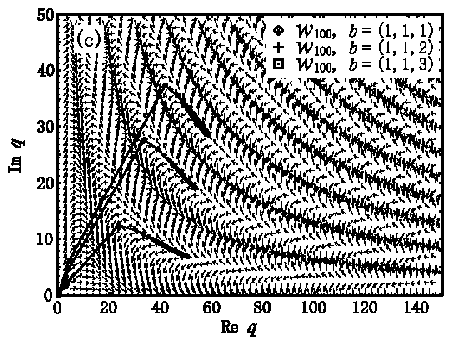}{0.585}{0.975}
\caption{\label{fig_obs2_no1}
(a), (b)
Trajectories of $w_j$'s for $j=$ 5, 10, and 15.
The real and imaginary parts of the trajectories are
displayed in (a) and (b) respectively.
The symbolic sequence of $w_j$ is given by
$\ldots\,\,O\,\,O\, .\,\,b\,\,b\,\ldots\,b\,\,O\,\,O\,\,\ldots$,
where the symbol $b$ is given in (b) and
the length of $b\,\,b\ldots b$ is $2j$.
The real and imaginary parts of each trajectory have
the maximal and minimal amplitudes respectively,
at time steps $j-1$ and $j$.
That is, the momentum almost vanishes at time step $j-1$.
(c) Trajectories of $w_j$'s with the same symbolic sequences 
as in (a) and (b) 
but for $j=100$ and $b=(1,1,\nu)$ with $\nu =1,2,3$. 
The vector field of $V^\prime (q)$ is superposed 
with the length normalized. 
In (a),(b), and (c), solid lines connecting 
points of trajectories are the guides for the eye.
}
\end{figure}

This observation is exemplified in Fig.~\ref{fig_obs2_no1}. 
In the case that the length of $\,b\,\,b\ldots b\,$ in
(\ref{symbolic_sequence_in_property3}) is given by $2j+1$ for $w_j$,
(\ref{turning_pts_in_property3_a}) and (\ref{turning_pts_in_property3_b})
hold in the range of $0<k\le j$, and
(\ref{turning_pts_in_property3_c}) and (\ref{turning_pts_in_property3_d})
hold in the range of $j<k\le 2j$.
In this case, the momentum $p_k(w_j)$ at $k=j$
is quite small, but does not vanish.

We conjecture that the $q$ component of the turning point,
$q_j(w_j)$, 
diverges with the length of $b\,\,b\ldots b$, i.e.,
the following relation holds:
\begin{eqnarray}\label{divergence_in_property3}
\lim_{j\rightarrow +\infty}q_j(w_j) = (x\!\cdot\!\infty,\,\,0),
\end{eqnarray}

\noindent
where $x$ is the sign of the infinity, which is given by the member of
the symbol $b=(x,y,\nu)$.
This conjecture is based on the following observation.

\begin{figure}
\outputfig{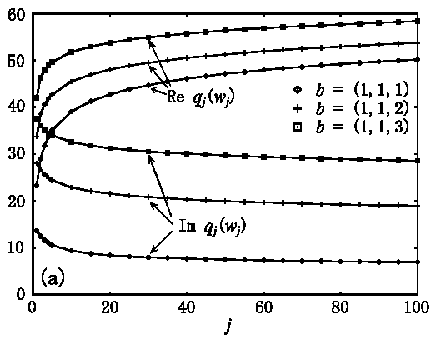}{0.6}{1.0}
\outputfig{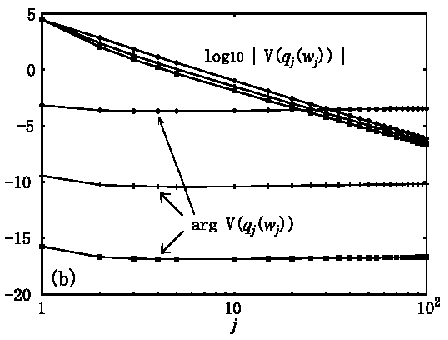}{0.6}{1.0}
\caption{\label{fig_obs2_no2}
(a)
The locations of the turning points, $q_j(w_j)$,
for the trajectories of $w_j$'s ranging from  $j=1$ to 100.
The symbolic sequences of $w_j$'s are the same as
is given in the caption of Fig. \ref{fig_obs2_no1}.
(b) 
The magnitude and phase of the potential function $V(q)$
at the turning point $q_j(w_j)$ for ranging from $j=1$ to 100.
The phase, $-2\gamma \hbox{Re}\,q_j\cdot\hbox{Im}\,q_j$,
is plotted without taking mod $2\pi$.
In (a) and (b), solid lines are the guides for the eye.
}
\end{figure}

Fig.~\ref{fig_obs2_no2} shows the locations of
the turning points $q_j(w_j)$ ranging from $j=1$ to 100,
and the magnitudes and phases of the potential function $V(q)$
at the turning points.
It can be seen that
as $j$ increases, the magnitude of $V(q_j(w_j))$
decreases to zero algebraically, and
the phase of $V(q_j(w_j))$ approaches some constant.
That is, $V(q_j(w_j))\approx\alpha j^{-\beta}e^{{\rm i}\theta}$ 
for large $j$'s,
where $\alpha ,\beta$ (both positive), and $\theta$ depend
on a symbol $b$ which consists of a consecutive part
$b\,b\ldots b$ in a symbolic sequence of $w_j$.
If this approximation holds for any large $j$,
solving the equation 
$V(q_j(w_j))=\alpha j^{-\beta}e^{{\rm i}\theta}$,
one obtains the solution:
\begin{subequations}\label{tp_on_uv_cdnt}
\begin{eqnarray}
u_j(w_j) &=& 
\left[ \beta\log j - \log (\alpha /k) \right] /(2\gamma ), \\
v_j(w_j) &=& -\theta /(2\gamma ),
\end{eqnarray}
\end{subequations}

\noindent
where $(u_j(w_j),v_j(w_j))$ is the location of $q_j(w_j)$
on the $(u,v)$ coordinate defined in (\ref{definition_of_hyp_cdn}).
This solution suggests that
$\hbox{Re}\,q_j(w_j)$ diverges and
$\hbox{Im}\,q_j(w_j)$ vanishes as $j\rightarrow +\infty$.
Moreover, according to the Observation 2,
$q_j(w_j)$ and the $q$ component of $U(b)$ are 
included in the same quadrant of the $q$ plane. 
Therefore the relation (\ref{divergence_in_property3}) is obtained.
The justification of this relation
needs further investigation of classical dynamics,
and we hope to report the result of this issue elsewhere.

We have shown that there are two types of behaviors exhibited
by homoclinic orbits.
In our numerical computations,
the behavior of any homoclinic orbits
can be understood by the combinations of only two types of motions,
one of which is 
the flipping motions almost along the axes $\hbox{Re}\,q = \pm\hbox{Im}\,q$,
and the other of which is the motions almost along
the contour curves of the $v$ component in the $(u,v)$ coordinate.
Which type of motion occurs in the process
from $(q_k,p_k)$ to $(q_{k+1},p_{k+1})$
along a single homoclinic trajectory depends on whether
the neighbouring symbols in a symbolic sequence, $a_k$ and $a_{k+1}$,
are the same (the latter type) or not (the former type).
The former type of motion is characterized by the Observation 1,
and the latter one by the Observation 2.

\subsection{Estimation of Imaginary Parts of Actions}
\label{title_estimation_of_imaginary_actions}

The imaginary parts of actions gained by the orbits of homoclinic
points 
are estimated from the symbolic sequences assigned to the points.
We first 
consider 
the homoclinic points appearing in the Observations 1 and 2,
and the estimations of imaginary parts of actions
for these cases are
presented as the Observations 3 and 4 respectively.
Then using the latter two observations, we estimate
the imaginary part of action for any homoclinic trajectory.
The Observation 3 says that the imaginary part of action diverges linearly
as $\nu\rightarrow +\infty$, where $\nu$ is a member of a symbol
$(x,y,\nu)$ in a symbolic sequence.
The Observation 4 says that the amount of
the imaginary part of action is bounded
even if 
the length of a consecutive part $b\,\,b\ldots b\,\,(b\in {\cal S})$
in a symbolic sequence tends to infinity.
In particular, we observed that the itinerary described by
$b\,\,b\ldots b$ gains little imaginary part of action
compared to the other itineraries in a trajectory.
This means that the homoclinic orbits appearing
in the Observation 4 can play a semiclassically significant role.
The Observations 3 and 4 are also entirely based on our numerical
computations, 
and so far, we have no mathematical proof for these observations.

For the trajectory of
any 
homoclinic point $w$,
here we consider the following imaginary part of action:
\begin{eqnarray}
s(w) & = & \sum_{k=1}^{+\infty}\hbox{Im}
\left[ T(p_{k-1}(w)) - V(q_k(w)) \right] ,
\label{def_of_classical_action}
\end{eqnarray}

\noindent
where $(q_k(w),p_k(w))$
denotes
the $k$ step iteration of $w$.
The sum in the r.h.s.
converges due to the exponential convergence of
the trajectory to the origin $(0,0)$,
and it is the long-time limit of
$\hbox{Im}\,S_n$ which is given in
(\ref{Van_Vleck_formula_d}).
In the definition of $s(w)$,
we only take account of the contributions from the forward trajectories,
since semiclassical wave functions in the time domain
are determined by them.
We do not define $s(w)$ as
the long-time limit of $\hbox{Im}\,{\tilde S}_n$
which is given in (\ref{Van_Vleck_formula_c}),
since it includes an additional term
$S_0$ which depends
only on the choice of an incident wave packet,
not on classical dynamics.
First we estimate $s(w)$
for the homoclinic points appearing in the Observation 1. \\

\noindent
Observation 3. 
Let $\left\{ w_1,w_2,w_3,\ldots\right\}$
be a set of homoclinic points
which appears in the Observation 1.
Then for any integer $n\ge 1$, the following relation holds:
\begin{eqnarray}\label{relation_in_obs3}
\lim_{\nu\rightarrow +\infty}
\left[\gamma /(2\nu\pi)\right]s( f ^{-n}(w_\nu))
&=& xy.
\end{eqnarray} \\

\begin{figure}
\outputfig{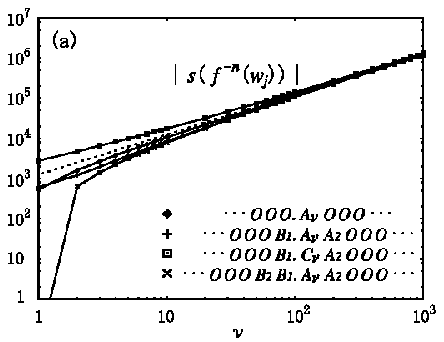}{0.585}{0.975}
\outputfig{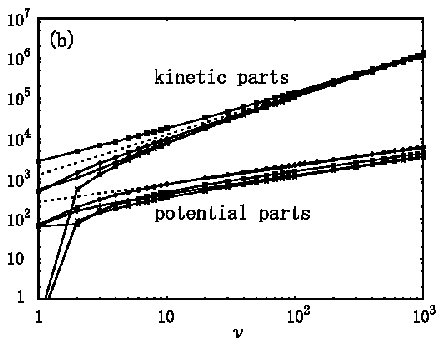}{0.585}{0.975}
\outputfig{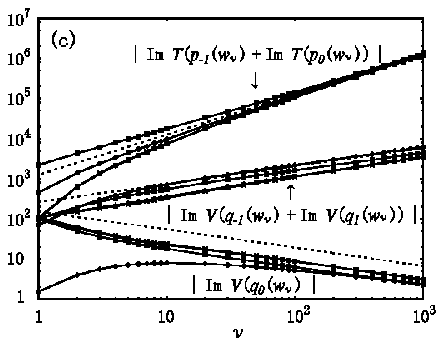}{0.585}{0.975}
\caption{\label{fig_Rule6}
(a)
The absolute values of
the imaginary parts of actions, $|s( f ^{-n}(w_{\nu}))|$,
for $n=4$ and $\nu = 1$ to $1000$.
The symbolic sequences of $w_{\nu}$'s are the same as in 
Fig. \ref{fig_obs1_no2}.
The dotted line represents $2\nu\pi /\gamma$.
(b)
The absolute values of
the kinetic parts and the potential ones
of $s( f ^{-n}(w_{\nu}))$.
The upper dotted line and the lower dotted curve represent
$2\nu\pi$ and $2f(\sqrt{2\nu\pi /\gamma})$ respectively,
where $f(x)=x/(2\sqrt{\gamma \log x})$.
(c)
$|\hbox{Im}\,T(p_{-1}(w_{\nu}))+\hbox{Im}\,T(p_0(w_{\nu}))|$,
$|\hbox{Im}\,V(q_{-1}(w_{\nu}))+\hbox{Im}\,V(q_1(w_{\nu}))|$, and
$|\hbox{Im}\,V(q_0(w_{\nu}))|$.
The top dotted line and middle dotted curve are the same as in (b), 
and the bottom dotted curve represents
$\{2\gamma^2 f(\sqrt{2\nu\pi /\gamma})\}^{-1}$.
In each figure, the solid lines are guides for the eye.
} 
\end{figure}

\noindent
Fig.~\ref{fig_Rule6}(a) shows the magnitudes of $s( f ^{-n}(w_j))$ for
$n=4$ 
and $\nu=1$ to $1000$.
It can be seen that $|s( f ^{-n}(w_{\nu}))|\approx 2\nu\pi /\gamma$
for large $\nu$'s. 
The condition $n\ge 1$ in (\ref{relation_in_obs3}) is imposed to
always 
take account of the contributions from
the flipping motions from $q_{-1}(w_{\nu})$ to $q_0(w_{\nu})$
displayed in Fig.~\ref{fig_obs1_no1}.

In the following, we explain the relation (\ref{relation_in_obs3}),
assuming that the Observation 1 holds, and that
for the homoclinic points appearing in the Observation 1,
$q_k(w_{\nu})$ (resp. $q_{-k}(w_{\nu})$) for $k=0,1,2,\ldots$ diverges
much faster than $q_{k+1}(w_{\nu})$ (resp. $q_{-(k+1)}(w_{\nu})$)
as $\nu\rightarrow +\infty$.
Fig.~\ref{fig_obs1_no3} suggests that
the second assumption is valid for $k=0$ and $1$.
For the other $k$'s, it has not been found numerically
whether the assumption is valid or not, since
$q_k(w_{\nu})$ and $q_{-k}(w_{\nu})$ for $k\ge 2$ remain to be immediate
values even for $\nu\approx 1000$, 
so that the numerical computation needs too high accuracy 
to make clear the asymptotic behaviors of $|q_k|$ and $|q_{-k}|$ 
with sufficiently large magnitudes. 
However, the exponential dependence of $V^{\prime}(q)$ on $u$ and $v$
shown in (\ref{estimations_of_amplitude_and_phase}) means that
the large difference in $q_k(w_j)$ (resp. $q_{-k)}(w_j)$) results from
the slight difference in $q_{k+1}(w_j)$ (resp. $q_{-(k+1)}(w_j)$) 
by the map $ f ^{-1}$ (resp. $ f $),
so that the assumption for $k\ge 2$ 
is expected to be valid.

The relation (\ref{relation_in_obs3}) is explained as follows.
In the kinetic part of $s( f ^{-n}(w_{\nu}))$, i.e.,
$\Sigma_{k=-n+1}^{+\infty}\hbox{Im}\,T(p_{k-1}(w_{\nu}))$,
each term is written as
\begin{eqnarray}
\hbox{Im}\,T(p_{k-1}(w_{\nu})) &=&
\left[\hbox{Re}\,p_{k-1}(w_{\nu})\right]
\left[\hbox{Im}\,p_{k-1}(w_{\nu})\right] .
\ \ \ \ \label{imaginary_kinetic}
\end{eqnarray}

\noindent
Due to the assumptions we put,
the following inequalities hold for $k\ne 0,-1$ and
for large $\nu$'s:
\begin{subequations}\label{difference_of_order_p}
\begin{eqnarray}
|\hbox{Re}\,p_k(w_{\nu})| &\ll&
|\hbox{Re}\,p_{-1}(w_{\nu})|,|\hbox{Re}\,p_0(w_{\nu})|,
\label{inequality_of_Rep} \\
|\hbox{Im}\,p_k(w_{\nu})| &\ll&
|\hbox{Im}\,p_{-1}(w_{\nu})|,|\hbox{Im}\,p_0(w_{\nu})|.
\label{inequailty_of_Imp}
\end{eqnarray}
\end{subequations}

\noindent
Then the kinetic part of $s( f ^{-n}(w_\nu))$ is dominated by the terms
$\hbox{Im}\,T(p_{-1}(w_{\nu}))$ and
$\hbox{Im}\,T(p_0(w_{\nu}))$ due to (\ref{imaginary_kinetic}).
Since the Observation 1 says that
the quantities in the r.h.s. of the above inequalities
have ${\nu}^{1/2}$ dependences on large $\nu$'s,
$\hbox{Im}\,T(p_{-1}(w_{\nu}))$ and $\hbox{Im}\,T(p_0(w_{\nu}))$
have linear dependences on large $\nu$'s
due to (\ref{imaginary_kinetic}).
Therefore the kinetic part of $s( f ^{-n}(w_\nu))$
is expected to has a linear dependence on large $\nu$'s.
Fig.~\ref{fig_Rule6} shows that
the kinetic part of $s( f ^{-n}(w_\nu ))$ is actually dominated by
$\hbox{Im}\,T(p_{-1}(w_{\nu}))$ and $\hbox{Im}\,T(p_0(w_{\nu}))$,
and has a linear dependence on large $\nu$'s.

In the potential part of $s( f ^{-n}(w_\nu))$, i.e.,
$\Sigma_{k=-n+1}^{+\infty}\hbox{Im}\,V(q_k(w_{\nu}))$,
each term is written as
\begin{eqnarray}
\hbox{Im}\,V(q_k(w_{\nu})) &=& \hbox{Im}\left[
\frac{q_{k+1}(w_{\nu})-2q_k(w_{\nu})+q_{k-1}(w_{\nu})}
{2\gamma q_k(w_{\nu})}\right] ,
\nonumber \\
& & \label{special_relation_of_V}
\end{eqnarray}

\noindent 
by incorporating the classical equations of motions
given in (\ref{classical_equations}) with the relation
$V^{\prime}(q)=-2\gamma qV(q)$ satisfied by our potential function.
A simple calculation of the r.h.s. of
(\ref{special_relation_of_V}) yields an inequality:
\begin{eqnarray}
|\hbox{Im}\,V(q_k(w_\nu ))| & \le &
\frac{|q_{k+1}(w_\nu )|+|q_{k-1}(w_\nu )|}{2\gamma |q_k(w_\nu )|} .
\label{special_inequality_of_V}
\end{eqnarray}

Based on the assumptions imposed here,
we can present the same discussion as that
below (\ref{asm_relation_in_obs1}) (note that the assumptions here
are stronger than those imposed there).
As a result, one obtains the relations
$|q_1(w_\nu )|,|q_{-1}(w_\nu )|\approx 
\sqrt{\gamma^{-1}\log |q_0(w_\nu )|}$ 
for large $\nu$.  
In a similar way, one obtains the relations
$|q_{\pm (k+1)}(w_\nu )|\approx \sqrt{\gamma^{-1}\log |q_{\pm k}(w_\nu ) |}$
for large $\nu$'s and $k=1,2,3,\ldots$.
By using these relations,
the r.h.s. of (\ref{special_inequality_of_V}) is approximated by
\begin{subequations}\label{estimation1_im_pot}
\begin{eqnarray}
\left[ 2\gamma^2 f(|q_k(w_{\nu})|)\right] ^{-1}
\quad &\hbox{for}& \quad k=0,
\label{estimation1_im_pot_a}\\
f(|q_{k-1}(w_{\nu})|)
\qquad &\hbox{for}& \quad k > 0,
\label{estimation1_im_pot_b}\\
f(|q_{k+1}(w_{\nu})|)
\qquad &\hbox{for}& \quad k < 0,
\label{estimation1_im_pot_c}
\end{eqnarray}
\end{subequations}

\noindent
where $f(x) = x/( 2\sqrt{\gamma\log x})$.
Here we used an approximation
$|q| + \sqrt{\gamma^{-1}\log |q|} \approx |q|$ for large $|q|$.

Since $|q_0(w_\nu )|$ is approximated as $\sqrt{2\nu\pi /\gamma}$
for large $\nu$'s according to the Observation 1,
the r.h.s. of (\ref{special_inequality_of_V}) is approximated by
\begin{subequations}\label{estimation2_im_pot}
\begin{eqnarray}
\left[ 2\gamma^2 f\left(\sqrt{2\nu\pi /\gamma}\right)\right] ^{-1}
\qquad\quad\! &\hbox{for}& k=0,
\ \ \ \ \ \ \label{estimation2_im_pot_a}\\
f\left(\sqrt{2\nu\pi /\gamma}\right) \qquad\qquad\!
&\hbox{for}& k=\pm 1,
\ \ \ \ \ \ \label{estimation2_im_pot_b}\\
f\left(\sqrt{\log^{\prime} (\log^{\prime} 
(\ldots \log^{\prime} (\log^{\prime}
(2\pi\nu /\gamma) )\ldots ))} \right)
\,&\hbox{for}& |k|\ge 2,
\ \ \ \ \ \ \label{estimation2_im_pot_c}
\end{eqnarray}
\end{subequations}

\noindent
where $\log^{\prime}x = (2\gamma )^{-1}\log x$, and
the argument of the square root in (\ref{estimation2_im_pot_c})
is a $|k|-1$ fold logarithm of $2\pi\nu / \gamma$.

Since $f(x)$ in (\ref{estimation2_im_pot}) is monotonically increasing
for large $x$, one can expect that
the potential part of $s( f ^{-n}(w_\nu ))$ for large $\nu$'s
is dominated by the terms
$\hbox{Im}\,V(q_{-1}(w_{\nu}))$ and $\hbox{Im}\,V(q_1(w_{\nu}))$.
More precisely, from (\ref{estimation2_im_pot_b}),
the potential part of $s( f ^n(w_\nu ))$
is expected to be approximated by $2f(\sqrt{2\nu\pi /\gamma})$
for large $\nu$'s. 
And also, from (\ref{estimation2_im_pot_a}),
$\hbox{Im}\,V(q_0(w_{\nu}))$ is expected to vanish as 
$\nu\rightarrow +\infty$. 
Fig.~\ref{fig_Rule6} shows that
the potential part of $s( f ^{-n}(w_\nu ))$ is actually dominated by
the terms $\hbox{Im}\,V(q_{-1}(w_{\nu}))$ and $\hbox{Im}\,V(q_1(w_{\nu}))$
for large $\nu$'s, 
and the asymptotic behavior of the potential part for large $\nu$'s
is described by $2f(\sqrt{2\nu\pi /\gamma})$.
It is also shown that
$\hbox{Im}\,V(q_0(w_{\nu}))$ tends to vanish as $\nu$ increases.

Consequently, 
due to the $\nu$ dependencies of 
kinetic and potential parts as seen above,
$s( f ^{-n}(w_\nu ))$ for large $\nu$'s
is dominated by the kinetic part,
which has a linear dependence of large $\nu$'s.
Therefore
the following estimation is obtained for large $\nu$'s:
\begin{eqnarray}
s( f ^{-n}(w_{\nu})) &\approx& 
\hbox{Im}\,T(p_{-1}(w_{\nu})) + \hbox{Im}\,T(p_0(w_{\nu}))
\nonumber \\
&=& \hbox{Re}\,p_{-1}(w_{\nu})\cdot\hbox{Im}\,p_{-1}(w_{\nu})
\nonumber \\
& & + \hbox{Re}\,p_0 (w_{\nu})\cdot\hbox{Im}\,p_0 (w_{\nu})
\nonumber \\
&\approx& (2\nu\pi /\gamma)xy.
\label{rough_estimation_of_ImS}
\end{eqnarray}

\noindent
In the last approximation, the relations
in (\ref{estimation_of_q_and_p_in_obs1})
are used.

Next we estimate $s(w)$
for the homoclinic points appearing in the Observation 2. \\

\noindent 
Observation 4. 
Let $\left\{ w_1,w_2,w_3,\ldots\right\}$
be a set of homoclinic points
which appears in the Observation 2.
Then for any integer $n$,
a sequence of imaginary parts of actions,
\begin{eqnarray}\label{obs4_seq_im_act}
s( f ^{-n}(w_1))\,\,\,s( f ^{-n}(w_2))\,\,\,
s( f ^{-n}(w_3))\ldots ,
\end{eqnarray}
is bounded. \\

\begin{figure}
\outputfig{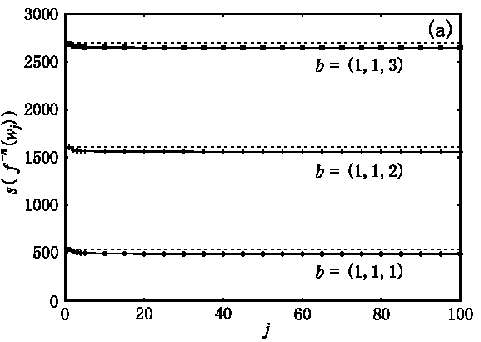}{0.6}{1.0}
\outputfig{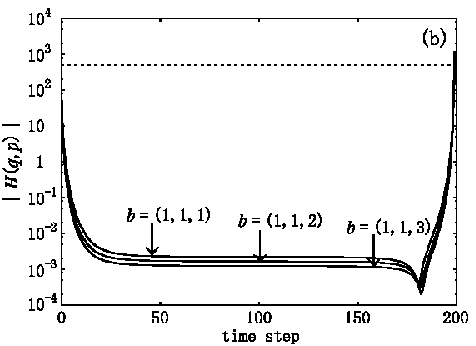}{0.6}{1.0}
\outputfig{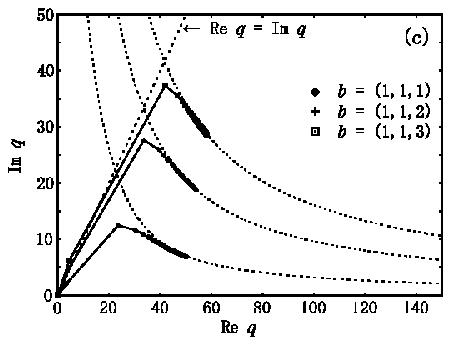}{0.6}{1.0}
\caption{\label{fig_obs4}
(a) 
The imaginary actions $s( f ^{-n}(w_j))$
plotted for $n=4$ and $j=1$ to 100.
In (a),(b), and (c), 
the symbolic sequence of $w_j$ is given by
$\ldots O\,O\, .\,\,b\,\,b\ldots \,b\,\,O\,O\,\ldots$,
where $b=(1,1,\nu)$ with $\nu =1,2,3$,
and the length of $b\,\,b\ldots \,b$ is given by $2j$.
The solid lines indicate the values of the imaginary parts of actions
in the case that the length of $b\,\,b\ldots b$ is one.
(b)
The magnitudes of the Hamiltonian $H(q,p)= p^2/2 + ke^{-\gamma q^2}$
(solid curves) 
along the forward trajectories of $w_j$'s for $j=100$.
The dotted line represents the magnitude of the Hamiltonian
at the origin $(0,0)$.
(c)
The homoclinic trajectories already shown in Fig. \ref{fig_obs2_no1}(c), 
and integrable trajectories (dotted curves) which satisfy
$H(q,p)=0$ and connect two infinities of the $q$ plane,
$(\hbox{Re}\,q,\hbox{Im}\,q)=(+\infty ,0)$ and $(0,+\infty )$.
The dotted line sprouting from the origin
represents $\hbox{Re}\,q=\hbox{Im}\,q$.
The solid lines are the guides for the eye.
}
\end{figure}

\noindent
Fig.~\ref{fig_obs4}(a) shows $s( f ^{-n}(w_j))$
for $n=4$ and $j=1$ to 100.
It can be seen that
$s( f ^{-n}(w_j))$ $(j=1,2,3,\ldots )$
does not deviate largely from $s( f ^{-n}(w))$ 
for a homoclinic point $w$ 
which has a symbolic sequence:
\begin{eqnarray}
\ldots O\,\,O\, . \,b\,\,\,O\,\,O\ldots .
\end{eqnarray} 
In the case of Fig.~\ref{fig_obs4}(a),
$s( f ^{-n}(w_j))$ $(j=1,2,3,\ldots )$ is mainly gained by 
the flippings of trajectories 
between phase-space components $U(b)$ and $U(O)$
($O=(0,0,0)$). 
Though not displayed here, 
also in the case of the symbolic sequence:
\begin{eqnarray}
\ldots a_{-2}\,a_{-1}\, .\,b\,\,\,b\ldots b
\,\,\,a_0\,\,a_1\ldots 
\end{eqnarray}
with $\ldots a_{-2}\,a_{-1}\ne \ldots O\,\,O$ 
or $a_0\,\,a_1\ldots \ne O\,\,O\ldots$, 
the imaginary part of action 
does not deviate largely from that in the case of the sequence:
\begin{eqnarray}
\ldots a_{-2}\,a_{-1}\, .\,b\,\,\,a_0\,\,a_1\ldots .
\end{eqnarray}

Therefore the imaginary parts of actions 
are mainly gained by phase-space itineraries 
described by sequences other than $b\,\,b\ldots b$. 
This means that 
itineraries described by $b\,\,b\ldots b$ can be
semiclassically significant in the tunneling process,
since smaller imaginary parts of actions yield
larger semiclassical amplitudes.
As in the Observation 2,
the same statement as the Observation 4 holds
when the length of the sequence $b\,\,b\ldots b$
is given by $2j+1$ for a homoclinic point $w_j$.

We discuss why 
the sequence (\ref{obs4_seq_im_act}) is bounded. 
Fig.~\ref{fig_obs4}(c) shows the same 
homoclinic trajectories as in Fig.~\ref{fig_obs2_no1}(c), 
and integrable trajectories of a Hamiltonian 
$H(q,p) = p^2/2 + ke^{-\gamma q^2}$. 
The integrable trajectories are chosen so that
they satisfy $H(q,p)=0$ and 
connect two infinities on the $q$ plane,
$(\hbox{Re}\,q, \hbox{Im}\,q) = (+\infty , 0)$ and $(0, +\infty)$.
It can be seen that the homoclinic trajectories
are along the integrable trajectories 
at intermediate time steps.
In fact, as shown in Fig.~\ref{fig_obs4}(b), 
$H(q,p)$ for the homoclinic trajectories almost vanishes 
at these time steps.

The previous 
Fig.~\ref{fig_obs2_no2}(b) implies that
the potential function at the turning points,
$V(q_{j}(w_j))$ $\left(\approx V(q_{j-1}(w_j))\right)$,
vanishes as $j\rightarrow +\infty$.
Then it is expected that
even for sufficiently large $j$'s, $H(q,p)\approx 0$ 
when the trajectories of $w_j$'s are around their turning points,
and that the trajectories 
are approximated
by the integrable trajectories satisfying $H(q,p)=0$.

If the above expectation is correct,
the imaginary parts of actions gained by the trajectories of $w_j$'s
described by a sequence $b\,\,b\ldots b$
can be 
evaluated 
by the integrable trajectories.
For our present parameter values $k=500$ and $\gamma =0.005$,
the magnitudes of the imaginary parts of actions are estimated
to be less than about 900 
(see the Appendix II). 
This value is an over-estimated one because homoclinic trajectories
and integrable ones do not agree very well around
the axis $\hbox{Re} q=\hbox{Im} q$ from where
the actions along integrable trajectories are integrated.

By making use of 
the Observations 3 and 4, we estimate
the imaginary parts of actions
gained by the orbits of homoclinic points
from the symbolic sequences assigned to the points.
Let $w$ be a homoclinic point whose symbolic sequence
is given by 
\begin{eqnarray}\label{sym_seq_in_estim_formula}
\ldots a_{-2}\,a_{-1}\,\, .\,\, a_0\,\, a_1\,\,a_2\ldots ,
\end{eqnarray}

\noindent
where $a_k = (x_k,y_k,\nu_k)$ for $k\in {\mathbb Z}$.
We assume that $\nu_k$ is large if $a_k\ne (0,0,0)$.

In the case that 
the symbolic sequence in (\ref{sym_seq_in_estim_formula})
does not include 
a consecutive part, $b\,\,b\ldots b$,
we approximate $q_k(w)$ for any $k\in {\mathbb Z}$
according to the Observation 1 by
\begin{eqnarray}\label{approximation_of_q}
q_k(w) &\approx& (\nu_k\pi /\gamma)^{1/2}\,\,(x_k,y_k).
\end{eqnarray}

\noindent
For $a_k =(0,0,0)$, substituting $(x_k,y_k)=(0,0)$
into (\ref{approximation_of_q}),
we approximate $q_k(w)$ by $0$.
From
the relation $p_{k-1} = q_k - q_{k-1}$,
we approximate $p_{k-1}(w)$ by
\begin{eqnarray}\label{approximation_of_p}
p_{k-1}(w) & \approx & (\pi /\gamma)^{1/2}
(x_k{\nu_k}^{1/2}-x_{k-1}{\nu_{k-1}}^{1/2}, \nonumber \\
& & y_k{\nu_k}^{1/2}-y_{k-1}{\nu_{k-1}}^{1/2}).
\end{eqnarray}

\noindent
Then due to (\ref{imaginary_kinetic}),
$\hbox{Im}\,T(p_{k-1}(w))$ is approximated by
\begin{eqnarray}\label{estimation_of_imaginary_kin_enegy}
\hbox{Im}\,T(p_{k-1}(w)) & \approx & (\pi /\gamma)
(x_k{\nu_k}^{1/2}-x_{k-1}{\nu_{k-1}}^{1/2}) \nonumber \\
& & \times (y_k{\nu_k}^{1/2}-y_{k-1}{\nu_{k-1}}^{1/2}).
\end{eqnarray}

\noindent
Since the imaginary part of action gained at each time step
is dominated by the kinetic part, as discussed at
the Observation 3, 
$s(w)$ is estimated by
the sum over the terms in the r.h.s. of
(\ref{estimation_of_imaginary_kin_enegy}) for $k\ge 1$.

In the case that the symbolic sequence
in (\ref{sym_seq_in_estim_formula})
includes 
a consecutive part, $b\,\,b\ldots b$,
the imaginary part of action gained
along the itinerary described by $b\,\,b\ldots b$ is negligible
compared to that along the other part of the trajectory,
as discussed below the Observation 4.
Based on this fact,
we approximate the imaginary part of action
for $b\,\,b\ldots b$ 
by a null value.
This approximation allows us to evaluate
the imaginary part of action only by
the kinetic part 
also in the case of $b\,\,b\ldots b$. 
It is because 
the r.h.s. of (\ref{estimation_of_imaginary_kin_enegy}) is null
when $(x_{k-1},y_{k-1},\nu_{k-1})=(x_k,y_k,\nu_k)$
so that the imaginary part of action associated with $b\,\,b\ldots b$
is evaluated as a null value.
As a result, whether 
a consecutive part $b\,\,b\ldots b$ is 
included in a symbolic sequence or not,
the imaginary part of action is estimated 
only by the kinetic part of action.

\begin{figure}
\outputfig{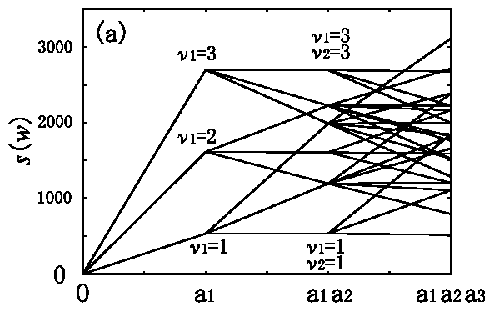}{0.6}{1.0}
\outputfig{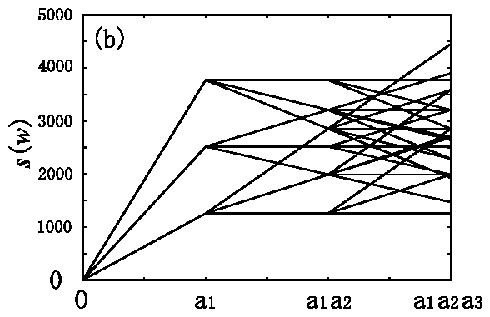}{0.6}{1.0}
\caption{\label{fig_tree_of_actions}
Imaginary parts of actions evaluated from
(a) actual trajectories of homoclinic points, and
(b) symbolic sequences assigned to the homoclinic points.
In each figure, 
the origin represents a null imaginary part of action
gained by the trajectory of 
a fixed point 
associated with a symbolic sequence
$\ldots O\,\,O\,\, .\,\,O\,\,O\,\,O\ldots\,\, 
\left( O=(0,0,0)\right)$. 
The first to the third columns labeled by
$a_1$,\,\, $a_1 a_2$, and $a_1 a_2 a_3$
show respectively the imaginary parts of actions
gained by the trajectories of the homoclinic points
associated with 
$\ldots O\,\,O\,\, .\,\,O\,\,O\,\,O\,\,a_1\,\,O\,\,O\ldots$,
$\ldots O\,\,O\,\, .\,\,O\,\,O\,\,O\,\,a_1\,\,a_2\,\,O\,\,O\ldots$, and
$\ldots O\,\,O\,\, .\,\,O\,\,O\,\,O\,\,a_1\,\,a_2\,\,a_3\,\,O\,\,O\ldots$,
where 
$a_1,a_2,a_3$ are the members of
$\{ (1,1,\nu )\,\,|\,\, \nu = 1,2,3  \}$.
}
\end{figure}

Finally, for the homoclinic point $w$ whose symbolic sequence
is given by (\ref{sym_seq_in_estim_formula}),
$s(w)$ is estimated as follows:
\begin{eqnarray}
s(w) &\approx&
\sum_{k=1}^{+\infty} \hbox{Im}\,T(p_{k-1}(w)) \nonumber \\
&\approx&
\frac{\pi}{\gamma}\sum_{k=1}^{+\infty}
(x_k{\nu}_k^{1/2} - x_{k-1}{\nu}_{k-1}^{1/2})
(y_k{\nu}_k^{1/2} - y_{k-1}{\nu}_{k-1}^{1/2}).
\nonumber \\
& & \label{derivation_of_estimation_formula}
\end{eqnarray}

\noindent
Fig.~\ref{fig_tree_of_actions} shows
the imaginary parts of actions
evaluated from actual trajectories of homoclinic points and
from symbolic sequences assigned to the homoclinic points.
The estimation (\ref{derivation_of_estimation_formula})
is based on the assumption that each $\nu_k$ is large if $a_k\ne (0,0,0)$,
however, as shown 
in the figure,
the estimation is still valid for small $\nu_k$.
This is because the approximation in (\ref{approximation_of_q})
are not so crude for small $\nu_k$, as is shown in
Fig.~\ref{fig_obs1_no2}(a).

In each column of Fig.~\ref{fig_tree_of_actions}(a), 
the smallest $s(w)$ is associated with 
a symbolic sequence 
$\ldots O\,\,O\, . \, O\,\,O\ldots$ or 
$\ldots O\,\,O\, . \, O\,\,O\,\,O\,\,b\,\,\,b\ldots b\,\,O\,\,O\ldots$ 
with $b=(1,1,1)$. 
The Appendix II shows 
such types of symbolic sequences, including the case 
that $b=(-1,-1,1)$, attain the smallest $s(w)$ 
in the whole candidates. 
In the previous 
Sec. \ref{title_reproduction_of_tunneling_amplitudes}, 
we evaluated the tunneling wave functions 
by the semiclassical candidate orbits 
described by these types of sequences,  
typical behavior of which was illustrated in 
Fig.~\ref{fig_orbits_and_symbols}(c).

\section{Conclusion and Discussion}
\label{title_discussion_and_conclusion}

\subsection{Conclusion}
\label{title_conclusion}

We have carried out complex semiclassical analysis for
the tunneling problem
of a simple scattering map which
creates chaotic dynamics in complex domain.
Although classical motions in real phase space are simple,
Tunneling wave functions exhibit a complicated pattern,
which is typically observed in chaotic systems.
The wave functions were reproduced semiclassically
in 
excellent
agreement with fully quantum calculations.
It enables us to interpret the creation of the complicated pattern
appearing in the tunneling regime.
Complex orbits contributing to the semiclassical
wave function are embedded in a hierarchical structure
of an initial-value set.
The hierarchical structure is a reflection of
the emergence of homoclinic tanglement in complex phase space,
i.e., the manifestation of complex-domain chaos.
On the basis of
symbolic dynamics constructed
in the complex domain,
phase space itineraries of tunneling orbits
were related with the amounts of imaginary parts of actions
gained by the orbits.
Incorporation of symbolic dynamics with
the complex semiclassical method has enabled us
to discuss quantitatively the competition among tunneling orbits,
and has elucidated the significant role of complex-domain chaos
in the tunneling process of non-integrable systems.

\subsection{Chaotic Tunneling}
\label{title_chaotic_tunneling}

We further discuss the role of complex-domain chaos
played in the semiclassical description of tunneling 
in non-integrable systems. 
In the present study, we adopted a time-domain approach 
of the complex semiclassical method. 
This approach is concerned with the real-time 
classical propagation, and has nothing to do with 
the instanton process. 
This means that real-domain paths are not connected 
to complex-domain paths, in other words, 
both the real domain and the other domain are invariant 
under the classical dynamics. 
Therefore all candidate orbits to describe tunneling 
are always exposed to complex-domain chaos, 
not to real-domain one. 
In this sense, it is natural to consider the role of 
the complex-domain chaos in our approach.

On our semiclassical framework, 
initial and final quantum states are identified with 
classical manifolds in complex phase space. 
The evolution of the manifolds is involved in 
the stretching and folding dynamics 
in the complex domain. 
The chain-like structure which we observed 
is nothing but the structure of 
the section of one backward evolved manifold, $f^{-n}({\cal F})$, 
cut by the other manifold ${\cal I}$. 
Our result here strongly suggests that 
the creation of the chain-like structure 
is only due to the emergence of complex-domain chaos, 
irrespective of the existence of real-domain chaos 
and also irrespective of the types of tunneling, i.e., 
whether energy-barrier tunneling or 
dynamical one
\cite{ShudoIkeda2}. 

The chaotic dynamics in phase space is created 
on the {\it Julia set}, which includes 
the complex homoclinic tanglement investigated here. 
The trajectories on this set are proven to be sufficient 
to describe tunneling in the case of the complex H\'enon map
\cite{ShudoIshiiIkeda}. 
It was numerically confirmed here that 
this statement is correct also in our case. 
Therefore,      
on the basis of our present study and 
Ref.~\cite{ShudoIshiiIkeda}, 
we would like to present the notion of ``chaotic tunneling'', 
which first appeared in 
Ref.~\cite{ShudoIkeda1}, 
as the tunneling in the presence of the Julia set.

In energy domain approaches
\cite{Frischat,Creagh,TakahashiYoshimotoIkeda,Brodier},
to our knowledge, 
the complex-domain chaos has not been used explicitly 
in semiclassical calculations. 
The significant role of the complex domain chaos played
in the time domain approach
should have the correspondence in the energy-domain ones.
However, the instanton concept, which is intrinsic to 
these approaches, 
makes it difficult to see such correspondence. 
The reason is that     
even when one takes full account of complex classical dynamics, 
the degree of freedom of the path deformation on the complex time plane 
often allows one to consider complicated classical processes 
in complex domain as the composition of real-domain chaotic processes 
and instanton-like ones
\cite{Creagh,TakahashiYoshimotoIkeda}. 
We would like to describe the tunneling phenomena 
in non-integrable systems in terms of the simple notion,
chaotic tunneling. 
Therefore the role of complex-domain chaos 
in the energy domain approaches 
is desired to be clarified in further studies.

\subsection{Discussion}
\label{title_discussion}

Finally, we here itemize several future problems where are necessary to
make our theory more self-contained:

1. We have constructed a partition of phase space
in terms of the phase part of the gradient of 
a potential function. 
A similar approach can be found
in the context of the study on a dynamical system of
an exponential map 
of one complex variable
\cite{Devaney},
where the boundaries of a partition correspond to
the contour curves of the phase part of
the exponential function.
Genericness of our approach should be
examined 
in further studies.

2. Reproducing tunneling wave functions,
we did not enter into details of the treatment of
the Stokes phenomenon.
Empirically, 
symbolic sequences which include members of the form
$(1,-1,\nu)$ or $(-1,1,\nu)$ with $\nu\in {\mathbb Z}$
should be excluded from the whole candidates.
In particular, according to such empirical rule, we have excluded
from the candidates
those trajectories which have almost null imaginary parts
of actions due to the cancellation
between the imaginary parts gained at individual time steps.
When the condition $q_{n+k} = q_{n-k-1}^{\ast}$ is satisfied
for any $k\ge 0$ with $n$ being fixed,
where the asterisk denotes the complex conjugate,
the imaginary parts of actions integrated over the whole time axis
become null. 
We observed numerically
that such condition is satisfied by the symbolic sequences
of homoclinic points such that the relation between symbols,
$(x_{n+k},y_{n+k},\nu_{n+k}) = (x_{n-k-1},-y_{n-k-1},\nu_{n-k-1})$,
holds for any $k\ge 0$ with $n$ being fixed.
In fact, there are an infinite number of symbolic sequences
satisfying such relation.
The criterion for whether
tunnneling orbits well approximated by
the homoclinic orbits described by such symbolic sequences
are semiclassically contributable or not
would be beyond our intuitive expectation
based on the amount of imaginary parts of actions
\cite{ShudoIkeda3}.
The criterion should be given only by a rigorous treatment of
the Stokes phenomenon.
The justification of our empirical rule mentioned above needs
the consideration of the intersection problem
of the Stokes curves, and we are now investigating this issue
\cite{OnishiShudo}.

3. In many non-integrable open systems with the condition that
$V^{\prime}(q)\rightarrow 0$ as $|q|\rightarrow +\infty$,
real-domain trajectories which diverge to infinity
are indifferent, i.e., have null Lyapunov exponents, in contrast to
the case of open systems with polynomial potential functions.
Because of that, in the former systems,
generaic properties of complex trajectories
exploring in the vicinity of real-domain asymptotic region
are not obvious, in spite of their semiclassical significant role
as has been seen in our present study.
The result of the investigation of this issue
will be reported elsewhere.

\begin{acknowledgements}
One of the authors(T.O.)  is grateful to A. Tanaka for stimulating
discussions. 
\end{acknowledgements}

\appendix

\section{Order of Symbolic Sequences
according to the Amounts of Imaginary Parts of Actions}

In this appendix, we give an order of symbolic sequences
according to the amounts of imaginary parts of actions
estimated by the formula (\ref{estimation_formula_of_ImS}).
We consider a set of symbolic sequences:
\begin{eqnarray}
\Sigma &=& \{ a_0\,\,a_1\ldots a_n\,\,O\,\,O\ldots \,\,|\,\,
n\ge 0, a_k\in {\cal S}^{\prime}
\,\,\hbox{for}\,\, k\ge 0 \} , \nonumber \\
& &
\end{eqnarray}

\noindent 
where
\begin{eqnarray}
{\cal S}^{\prime} &=& \{ (1,1,\nu ),(-1,-1,\nu)
\,\,|\,\,\nu\in {\mathbb N}\}
\cup \{ (0,0,0) \}. \ \ \ \
\end{eqnarray}

For any two members of $\Sigma$,
$\sigma$ and $\sigma^{\prime}$,
we introduce an equivalent relation `$\sim$' by
\begin{eqnarray}
\sigma \sim \sigma^{\prime} \quad &\Leftrightarrow& \quad
\tilde{s}(\sigma ) = \tilde{s}(\sigma^{\prime}),
\end{eqnarray}

\noindent
where $\tilde{s}(\sigma)$ denotes the imaginary part of action
estimated by (\ref{estimation_formula_of_ImS})
for the symbolic sequence $\sigma$.
For any two members of $\Sigma /\!\sim$,
$[\sigma ]$ and $[\sigma^{\prime}]$,
the order between them is defined by
\begin{eqnarray}
[\sigma ] < [\sigma^{\prime}] \quad \Leftrightarrow \quad
\tilde{s}(\sigma ) < \tilde{s}(\sigma^{\prime}),
\end{eqnarray}

It is easily checked that
$\tilde{s}(\sigma)\ge 0$ for any $\sigma\in\Sigma$
and the equality holds if and only if 
$\sigma = O\,\,O\,\,O\ldots $.
Then one obtains 
\begin{eqnarray}
[O\,\,O\,\,O\ldots ] < [\sigma ] \quad\Leftrightarrow\quad
\sigma \ne O\,\,O\,\,O\ldots .
\label{appendixI_a}
\end{eqnarray}

\noindent
For any $[\sigma]$ with $[\sigma ]\ne [O\,\,O\,\,O\ldots ]$,
the symbolic sequence $\sigma$ can be chosen so as to take a form:
\begin{eqnarray}
\sigma  & = & a_0\,\,a_1\ldots a_n\,\, O\,\,O\ldots\quad 
(n\ge 0), 
\label{appendixI_b1}
\end{eqnarray}

\noindent
which satisfies 
\begin{eqnarray}
a_n\ne O,\quad a_{k-1}\ne a_k\quad\hbox{for}\quad 0\le k\le n.
\label{appendixI_b2}
\end{eqnarray}

\noindent
For the symbolic sequence $\sigma$ in 
(\ref{appendixI_b1}) and (\ref{appendixI_b2}),
the following relations hold:
\begin{eqnarray}
& & [a_n\,\,O\,\,O\ldots ]<[a_{n-1}\,a_n\,\,O\,\,O\ldots ]< \ldots
\nonumber \\
& & \ldots < 
[a_0\,\,a_1\ldots a_n\,\,O\,\,O\ldots ] = [\sigma ].
\label{appendixI_c}
\end{eqnarray}

\noindent
Since, for any $\nu\in {\mathbb N}$,
\begin{eqnarray}
&   & \tilde{s}((1,1,\nu)\,\,O\,\,O\ldots )     \nonumber\\
& = & \tilde{s}((-1,-1,\nu)\,\,O\,\,O\ldots )   \nonumber\\
& < & \tilde{s}((-1,-1,\nu +1)\,\,O\,\,O\ldots )\nonumber\\
& = & \tilde{s}((1,1,\nu +1)\,\,O\,\,O\ldots ),
\end{eqnarray}

\noindent
either of the following relations holds:
\begin{subequations}\label{appendixI_d}
\begin{eqnarray}
\left[(1,1,1)\,\,O\,\,O\ldots\right] & < & [\sigma ] , \\
\left[(1,1,1)\,\,O\,\,O\ldots\right] & = & [\sigma ] .
\end{eqnarray}
\end{subequations}

Finally, from (\ref{appendixI_a}) and (\ref{appendixI_d}),
one obtains the following relations:
\begin{eqnarray}
& & [O\,\,O\,\,O\ldots ]<[(1,1,1)\,\,O\,\,O\ldots ]<[\sigma ],
\end{eqnarray}

\noindent
for $[\sigma ]$ with
$[\sigma ]\ne [O\,\,O\,\,O\ldots],\, [(1,1,1)\,\,O\,\,O\ldots ]$.
It is not difficult to check that
\begin{eqnarray}
&   & [(1,1,1)\,\,O\,\,O\ldots ] \nonumber \\
& = & \{b\,\,\,b\ldots b\,\,O\,\,O\ldots \,\,|\,\,
b = (1,1,1) \,\, \hbox{or} \,\, (-1,-1,1)\}.
\ \ \ \ \ \ \ \
\end{eqnarray}

\section{Imaginary Parts of Actions 
for Integrable Trajectories}

In this appendix, we evaluate the imaginary parts of actions
for integrable trajectories which satisfy $T(p)+V(q)=0$
and connect two infinities of the $q$ plane,
$(\hbox{Re}\,q,\hbox{Im}\,q)=(+\infty ,0)$ and $(0,+\infty)$.
These trajectories are included in the first quadrant
of the $q$ plane as shown in Fig.~\ref{fig_obs4}.
There are symmetric counterparts of the trajectories
in the other quadrants, and the application of the result here
to them is straightforward.

From the relation $T(p)+V(q)=0$, 
one obtains
\begin{eqnarray}\label{appendix_IIa}
p(q) &=& \pm\, {\rm i}\sqrt{2k}\,e^{-\gamma q^2/2}.
\end{eqnarray}

\noindent
Then the action $S(q,q^{\prime})$ defined by
\begin{eqnarray}\label{appendix_IIb}
S(q,q^{\prime}) & = &
\int_{q}^{q^{\prime}}\left[ T(p)-V(q)\right]\frac{dq}{p}, 
\end{eqnarray}
can be written as 
\begin{eqnarray}\label{appendix_IIc}
S(q,q^{\prime}) &=& \pm\, {\rm i}2\sqrt{2k}
\int_{q}^{q^{\prime}}e^{-\gamma q^2/2}dq.
\end{eqnarray}

We denote $q_0=(0,0)$, $q_{\infty}=(+\infty,0)$, and
$q_x=(x,x)$ with $x\ge 0$.
The action integrated along the real axis from $q_0$ to $q_{\infty}$
is evaluated immediately as
\begin{eqnarray}\label{appendix_IId}
S(q_0,q_{\infty}) &=& \pm\, {\rm i}2\sqrt{\pi k/\gamma}.
\end{eqnarray}
Let $l$ be one of the integrable trajectories on the $q$ plane
and $(x^{\prime},x^{\prime})$ be the intersection point between
$l$ and the axis $\hbox{Re}\, q = \hbox{Im}\, q$.
Deforming the integral path, the above action is represented as
\begin{eqnarray}\label{appendix_IIe}
S(q_0,q_{\infty}) &=& S(q_0,q_{x^{\prime}}) + S(q_{x^{\prime}},q_{\infty}),
\end{eqnarray}
where $S(q_0,q_{x^{\prime}})$ and $S(q_{x^{\prime}},q_{\infty})$
are integrated along the axis $\hbox{Re}\, q = \hbox{Im}\, q$ and
the path $l$ respectively.

$S(q_0,q_{x^{\prime}})$ is represented as
\begin{eqnarray}\label{appendix_IIf}
\pm\,2\sqrt{\pi k/\gamma }
\left\{ \left[ -C(w)+S(w)\right] 
+{\rm i}\left[  C(w)+S(w)\right]\right\},
\end{eqnarray}
where $w=\sqrt{2\gamma /\pi}x^{\prime}$ and
$C(w)$, $S(w)$ are the Fresnel's functions defined by
\begin{subequations}\label{appendix_IIg}
\begin{eqnarray}
C(w)=\int_0^w\cos (\pi t^2/2)dt,\\
S(w)=\int_0^w\sin (\pi t^2/2)dt.
\end{eqnarray}
\end{subequations}
From (\ref{appendix_IId}), (\ref{appendix_IIe}), 
and (\ref{appendix_IIf}), one obtains
\begin{eqnarray}\label{appendix_IIh}
\hbox{Im}\,\,S(q_{x^{\prime}},q_{\infty})
&=& \pm\,2\sqrt{\pi k/\gamma}\left[ 1-C(w)-S(w)\right] .
\end{eqnarray}

For our parameter values $k=500$ and $\gamma = 0.005$,
we observed that any integrable trajectory connecting infinities
$(+\infty ,0)$ and $(0,+\infty)$ on the $q$ plane
satisfies $w>1.0$. 
Since $0.6< C(w)+S(w) <1.4$ in this range of $w$,
we estimate the imaginary part of action as
\begin{equation}\label{appendix_IIi}
|\hbox{Im}\,\,S(q_{x^{\prime}},q_{\infty})|< 450.
\end{equation}
Since $C(w)$ and $S(w)$ converge to 1/2 as $w\rightarrow +\infty$,
$\hbox{Im}\,\,S(q_{x^{\prime}},q_{\infty})$ vanishes in this limit.

\end{document}